\documentclass[twocolumn]{aastex63}
\pdfoutput=1 
\usepackage{amsmath,amstext}
\usepackage[T1]{fontenc}
\usepackage{apjfonts}
\usepackage{hyperref}
\usepackage[all]{hypcap}
\usepackage{url}
\usepackage{xspace}
\usepackage{color}

\usepackage{savesym}
\savesymbol{tablenum}
\usepackage{siunitx}
\restoresymbol{SIX}{tablenum}

\usepackage{lineno}

\usepackage{booktabs}

\newcommand{\spider}{\textsc{Spider}\xspace}
\newcommand{\Spider}{\textsc{Spider}\xspace}
\newcommand{\planck}{\textit{Planck}\xspace}

\shorttitle{XFaster Pipeline}
\shortauthors{\Spider Collaboration}

\begin{document}

\title{The XFaster Power Spectrum and Likelihood Estimator for the Analysis of Cosmic Microwave Background Maps}

\newcommand\ANL{High Energy Physics Division, Argonne National Laboratory, Argonne, IL, USA 60439}
\newcommand\CWRU{Physics Department, Case Western Reserve University, 10900 Euclid Ave, Rockefeller Building, Cleveland, OH 44106, USA}
\newcommand\Cardiff{School of Physics and Astronomy, Cardiff University, The Parade, Cardiff, CF24 3AA, UK}
\newcommand\UBC{Department of Physics and Astronomy, University of British Columbia, 6224 Agricultural Road,
Vancouver, BC V6T 1Z1, Canada}
\newcommand\Princeton{Department of Physics, Princeton University, Jadwin Hall, Princeton, NJ 08544, USA}
\newcommand\Caltech{Division of Physics, Mathematics and Astronomy, California Institute of Technology, MS 367-17, 1200 E. California Blvd., Pasadena, CA 91125, USA}
\newcommand\JPL{Jet Propulsion Laboratory, Pasadena, CA 91109, USA}
\newcommand\CITA{Canadian Institute for Theoretical Astrophysics, University of Toronto, 60 St. George Street, Toronto, ON M5S 3H8, Canada}
\newcommand\ASU{School of Electrical, Computer, and Energy Engineering, Arizona State University, 650 E Tyler Mall, Tempe, AZ 85281, USA}
\newcommand\UKZN{School of Mathematics, Statistics and Computer Science, University of KwaZulu-Natal, Durban, South Africa}
\newcommand\NITP{National Institute for Theoretical Physics (NITheP), KwaZulu-Natal, South Africa}
\newcommand\Imperial{Blackett Laboratory, Imperial College London, SW7 2AZ, London, UK}
\newcommand\Stockholm{The Oskar Klein Centre for Cosmoparticle Physics, Department of Physics, Stockholm University, AlbaNova, SE-106 91 Stockholm, Sweden}
\newcommand\Oslo{Institute of Theoretical Astrophysics, University of Oslo, P.O. Box 1029 Blindern, NO-0315 Oslo, Norway}
\newcommand\TorontoDunlap{Dunlap Institute for Astronomy and Astrophysics, University of Toronto, 50 St George Street, Toronto, ON M5S 3H4 Canada}
\newcommand\Toronto{Department of Astronomy and Astrophysics, University of Toronto, 50 St George Street, Toronto, ON M5S 3H4 Canada}
\newcommand\UIUCP{Department of Physics, University of Illinois at Urbana-Champaign, 1110 W. Green Street, Urbana, IL 61801, USA}
\newcommand\UIUCA{Department of Astronomy, University of Illinois at Urbana-Champaign, 1002 W. Green Street, Urbana, IL 61801, USA}
\newcommand\NRAO{National Radio Astronomy Observatory, Charlottesville, NC 22903, USA}
\newcommand\Michigan{Department of Physics, University of Michigan, 450 Church Street, Ann  Arbor, MI 48109, USA}
\newcommand\TorontoP{Department of Physics, University of Toronto, 60 St George Street, Toronto, ON M5S 3H4 Canada}
\newcommand\Hopkins{Department of Physics and Astronomy, Johns Hopkins University, 3701 San Martin Drive, Baltimore, MD 21218 USA}
\newcommand\Goddard{NASA Goddard Space Flight Center, Code 665, Greenbelt, MD 20771, USA}
\newcommand\APC{APC, Univ. Paris Diderot, CNRS/IN2P3, CEA/Irfu, Obs de Paris, Sorbonne Paris Cit\'e, France}
\newcommand\PennState{Department of Astronomy and Astrophysics, Pennsylvania State University, 520 Davey Lab, University Park, PA 16802, USA}
\newcommand\NIST{National Institute of Standards and Technology, 325 Broadway Mailcode 817.03, Boulder, CO 80305, USA}
\newcommand\Stanford{Department of Physics, Stanford University, 382 Via Pueblo Mall, Stanford, CA 94305, USA}
\newcommand\SLAC{ SLAC National Accelerator Laboratory, 2575 Sand Hill Road, Menlo Park, CA 94025, USA}
\newcommand\PrincetonEngineering{Department of Mechanical and Aerospace Engineering, Princeton University, Engineering Quadrangle, Princeton, NJ 08544, USA}
\newcommand\Fermilab{Fermi National Accelerator Laboratory, P.O. Box 500, Batavia, IL 60510-5011, USA}
\newcommand\KICPChicago{Kavli Institute for Cosmological Physics, University of Chicago, 5640 S Ellis Avenue, Chicago, IL 60637 USA}
\newcommand\Orsay{Institut d'Astrophysique Spatiale, Orsay, France}
\newcommand\MPI{Max-Planck-Institute for Astronomy, Konigstuhl 17, 69117, Heidelberg, Germany}
\newcommand\LAIM{Laboratoire AIM, Paris-Saclay, CEA/IRFU/SAp - CNRS - Universit\'e Paris Diderot, 91191, Gif-sur-Yvette Cedex, France}
\newcommand\WUSTL{Department of Physics, Washington University in St. Louis, 1 Brookings Drive, St.  Louis, MO 63130, USA}
\newcommand\MCSS{McDonnell Center for the Space Sciences, Washington University in St. Louis, 1 Brookings Drive, St.  Louis, MO 63130, USA}
\newcommand\Austin{Department of Physics, University of Texas, 2515 Speedway, C1600, Austin, TX 78712, USA}
\newcommand\McGill{Department of Physics, McGill University, 3600 Rue University, Montreal, QC, H3A 2T8, Canada}
\newcommand\StewardObs{Steward Observatory, 933 North Cherry Avenue, Tucson, AZ, 85721, USA}
\newcommand\Shahid{Department of Physics, Shahid Beheshti University, 1983969411, Tehran Iran}

\author{ A.~E.~Gambrel }
\affiliation{\KICPChicago}

\author{ A.~S.~Rahlin }
\affiliation{\Fermilab}
\affiliation{\KICPChicago}

\author{ X.~Song }
\affiliation{\Princeton}

\author{ C.~R.~Contaldi }
\affiliation{\Imperial}

\author{ P.~A.~R.~Ade }
\affiliation{\Cardiff}

\author{ M.~Amiri }
\affiliation{\UBC}

\author{ S.~J.~Benton }
\affiliation{\Princeton}

\author{ A.~S.~Bergman }
\affiliation{\Princeton}

\author{ R.~Bihary }
\affiliation{\CWRU}

\author{ J.~J.~Bock }
\affiliation{\Caltech}
\affiliation{\JPL}

\author{ J.~R.~Bond }
\affiliation{\CITA}

\author{ J.~A.~Bonetti }
\affiliation{\JPL}

\author{ S.~A.~Bryan }
\affiliation{\ASU}

\author{ H.~C.~Chiang }
\affiliation{\McGill}
\affiliation{\UKZN}

\author{ A.~J.~Duivenvoorden }
\affiliation{\Princeton}
\affiliation{\Stockholm}

\author{ H.~K.~Eriksen }
\affiliation{\Oslo}

\author{ M.~Farhang }
\affiliation{\Shahid}
\affiliation{\CITA}
\affiliation{\Toronto}

\author{ J.~P.~Filippini }
\affiliation{\UIUCP}
\affiliation{\UIUCA}

\author{ A.~A.~Fraisse }
\affiliation{\Princeton}

\author{ K.~Freese }
\affiliation{\Austin}
\affiliation{\Stockholm}

\author{ M.~Galloway }
\affiliation{\Oslo}

\author{ N.~N.~Gandilo }
\affiliation{\StewardObs}

\author{ R.~Gualtieri}
\affiliation{\ANL}

\author{ J.~E.~Gudmundsson }
\affiliation{\Stockholm}

\author{ M.~Halpern }
\affiliation{\UBC}

\author{ J.~Hartley }
\affiliation{\TorontoP}

\author{ M.~Hasselfield }
\affiliation{\PennState}

\author{ G.~Hilton }
\affiliation{\NIST}

\author{ W.~Holmes }
\affiliation{\JPL}

\author{ V.~V.~Hristov }
\affiliation{\Caltech}

\author{ Z.~Huang }
\affiliation{\CITA}

\author{ K.~D.~Irwin }
\affiliation{\Stanford}
\affiliation{\SLAC}

\author{ W.~C.~Jones }
\affiliation{\Princeton}

\author{ A.~Karakci }
\affiliation{\Oslo}

\author{ C.~L.~Kuo }
\affiliation{\Stanford}

\author{ Z.~D.~Kermish }
\affiliation{\Princeton}

\author{ J.~S.-Y.~Leung }
\affiliation{\Toronto}
\affiliation{\TorontoDunlap}

\author{ S.~Li }
\affiliation{\Princeton}
\affiliation{\PrincetonEngineering}

\author{D.~S.~Y.~Mak}
\affiliation{\Imperial}

\author{ P.~V.~Mason }
\affiliation{\Caltech}

\author{ K.~Megerian }
\affiliation{\JPL}

\author{ L.~Moncelsi }
\affiliation{\Caltech}

\author{ T.~A.~Morford }
\affiliation{\Caltech}

\author{ J.~M.~Nagy }
\affiliation{\WUSTL}
\affiliation{\MCSS}

\author{ C.~B.~Netterfield }
\affiliation{\Toronto}
\affiliation{\TorontoP}

\author{ M.~Nolta }
\affiliation{\CITA}

\author{ R. O\rq Brient}
\affiliation{\JPL}

\author{ B.~Osherson }
\affiliation{\UIUCP}

\author{ I.~L.~Padilla }
\affiliation{\Toronto}
\affiliation{\Hopkins}

\author{ B.~Racine }
\affiliation{\Oslo}

\author{ C.~Reintsema }
\affiliation{\NIST}

\author{ J.~E.~Ruhl }
\affiliation{\CWRU}

\author{ T.~M.~Ruud }
\affiliation{\Oslo}

\author{ J.~A.~Shariff }
\affiliation{\CITA}

\author{ E.~C.~Shaw }
\affiliation{\UIUCP}

\author{ C.~Shiu }
\affiliation{\Princeton}

\author{ J.~D.~Soler }
\affiliation{\MPI}

\author{ A.~Trangsrud }
\affiliation{\Caltech}
\affiliation{\JPL}

\author{ C.~Tucker }
\affiliation{\Cardiff}

\author{ R.~S.~Tucker }
\affiliation{\Caltech}

\author{ A.~D.~Turner }
\affiliation{\JPL}

\author{ J.~F.~van~der~List }
\affiliation{\Princeton}

\author{ A.~C.~Weber }
\affiliation{\JPL}

\author{ I.~K.~Wehus }
\affiliation{\Oslo}

\author{ S.~Wen }
\affiliation{\CWRU}

\author{ D.~V.~Wiebe }
\affiliation{\UBC}

\author{ E.~Y.~Young }
\affiliation{\Stanford}
\affiliation{\SLAC}

\correspondingauthor{Anne Gambrel}
\email{agambrel@kicp.uchicago.edu}

\begin{abstract}
  We present the XFaster analysis package. XFaster is a fast, iterative angular power spectrum estimator based on a diagonal approximation to the quadratic Fisher matrix estimator.
  XFaster uses Monte Carlo simulations to compute noise biases and filter transfer functions and is thus a hybrid of both Monte Carlo and quadratic estimator methods.
  In contrast to conventional pseudo-$C_\ell$ based methods, the algorithm described here requires a minimal number of simulations, and does not require them to be precisely representative of the data to estimate accurate covariance matrices for the bandpowers.
  The formalism works with polarization-sensitive observations and also data sets with identical, partially overlapping, or independent survey regions.
  The method was first implemented for the analysis of BOOMERanG data \citep{Netterfield2002,xfaster_boomerang}, and also used as part of the \planck analysis \citep{xfaster_planck}.
  Here, we describe the full, publicly available analysis package, written in Python, as developed for the analysis of data from the 2015 flight of the \spider instrument \citep{bmode_paper}.
  The package includes extensions for self-consistently estimating null spectra and for estimating fits for Galactic foreground contributions.
  We show results from the extensive validation of XFaster using simulations, and its application to the \spider data set.
\end{abstract}

\section{Introduction}
Analysis of cosmic microwave background (CMB) observations generally
requires the compression of large, time-domain data sets into the much
smaller space of angular power spectra, $C_\ell$. This compression is
typically achieved in multiple stages. The first stage involves a
compression of the time-domain data into a higher
signal-to-noise map of the sky \citep{madcap}. The second
stage involves a compression from the map-domain to the angular power
spectrum \citep{bjk}. In order to achieve an unbiased, optimal, and
lossless compression, both these stages require a number of
assumptions to hold. The first compression relies on the assumption
that the residual between the time-domain data and a signal model is
distributed as a Gaussian variate. In practice, the noise is also assumed to be
stationary over sufficiently long timescales such that it can be
modeled efficiently in the Fourier domain over a useful range of frequencies. 
The second stage assumes both
signal and noise components of the map are Gaussian-distributed random
variates with known pixel-to-pixel covariances. In principle, if these
assumptions hold, optimal, lossless maps of the sky can be obtained
using a closed-form solution of the $\chi^2$ minimization problem, and
the maps can be compressed to a final set of $C_\ell$s through a
numerical maximization of a likelihood of the map.

In practice, a number of complications limit the validity of these
assumptions. This is particularly the case for ground-based or
sub-orbital observations, where scan strategies limit long-term stability compared to
space-based observations and introduce a number of systematics.
These include beam
asymmetries, atmospheric effects, instrumental noise, and thermal
instabilities. These systematics are difficult, or sometimes impossible, to
account for using covariances describing stationary, statistically
isotropic random variates. Instead, many observations are modeled
using end-to-end simulations where non-idealities can be included more
easily.
The simulations are used to calibrate templates of systematic
effects to be subtracted from the data, to calculate noise biases, and
to determine the distribution of estimated quantities. The presence of sky cuts, foregrounds, and
inhomogeneous pixel coverage can also be modeled easily in end-to-end
simulations. 
Simulation-based methods have been employed successfully in the analysis of
ground-based, suborbital, and space-based observations (\emph{e.g.},
\cite{Chiang_2010}, \cite{Netterfield2002}, \cite{planck18_params}).

One of the first methods to use simulations to estimate power spectra
is the MASTER formalism \citep{master}. In this approach, the full-sky
$C_\ell$s are estimated from the cut-sky
pseudo-$C_\ell$s, $\widetilde{C}_\ell$, by subtracting a noise bias and
dividing by a filter function, both of which are calibrated using
noise-only and signal-only ensembles of sky maps obtained by end-to-end
simulations of the time-domain data. The $\widetilde{C}_\ell$s are related
to the full-sky $C_\ell$s using coupling kernels that can be calculated
from the weighted cuts imposed on the sky maps.

As long as the noise and signal simulations are representative of the
data, this method gives an unbiased estimate of the bandpowers, and
their covariance is determined from the simulations.  However, it is
non-trivial to produce accurate signal and noise
simulations, and in general, the MASTER method requires iteration of
the simulated map ensembles to produce an accurate covariance.
Since map generation using end-to-end
simulations of the time-domain data tends to be the most
computationally expensive step in most CMB analysis pipelines, this
can be inefficient or intractable for modern data sets.

This paper details the XFaster method and demonstrates its
application on the \spider 2015 flight data. The XFaster method blends the
maximum-likelihood approach of \citet{bjk} with the MASTER
approach by introducing an approximate likelihood for the data that is
calibrated using simulation ensembles. There are a number of
advantages to this approach. The definition of an approximate likelihood
allows the use of a quadratic estimator to obtain the $C_\ell$s with a
simultaneous estimate of a Fisher matrix. The likelihood method can be
extended to include marginalization over additional signals, systematics, or
noise biases. Prior constraints are also easily included when a
likelihood is defined. Ancillary quantities such as bandpower window functions
can be calculated from the maximum-likelihood estimator. It also
reduces the number of simulations required to calibrate the noise and
filter biases to a minimum set of noise-only and signal-only
simulations which do not necessarily need to be fiducial matches to
the data. The introduction of an approximate likelihood can also be
used to define a higher-level likelihood for model parameters given
the data. This allows XFaster to be used in likelihood samplers to
obtain estimates of model parameters directly from the map-level data.

The XFaster method was originally conceived for the analysis of
data from the BOOMERanG flights \citep{Netterfield2002,xfaster_boomerang}. The application to
BOOMERanG data motivated the main development of the XFaster
pipeline, originally written as a Fortran90 package.
It was also applied to \planck data \citep{xfaster_planck}.
Here, we comprehensively review a revamped and extended version,
re-written as a versatile Python package, for the analysis of \spider data.
\spider is a balloon-borne polarimeter that was launched on January 1, 2015, from the NASA/NSF Long-Duration Balloon facility near McMurdo Station, Antarctica.
It mapped 2,480~square degrees at \SIlist{95;150}{\giga\hertz} during its 16.5-day flight.
Details relevant to the application of XFaster on this data set, as presented in \cite{bmode_paper}, are given throughout.
However, while this paper uses \spider simulations and data to demonstrate the functionality of the pipeline, the methods and publicly available code base are intended for general use for any CMB observatory.

The paper is organized as follows.
In Section~\ref{sec:algorithm}, we introduce the XFaster algorithm for parameter likelihoods and bandpowers.
In Section~\ref{sec:extensions}, we discuss extensions to the baseline algorithm: null tests and foregrounds.
In Section~\ref{sec:validation}, we show the results of running the pipeline on simulations.
Next in Section~\ref{sec:spiderdata}, we show results from running the pipeline on data from \spider's 2015 flight.
We discuss details of the public code base and its computational requirements in Section~\ref{sec:code}, and we conclude in Section~\ref{sec:conclusion}.

\section{The XFaster Algorithm}
\label{sec:algorithm}

\subsection{Likelihood Approximation}
When experiments observe only a fraction of the sky, or when a portion
of the sky is excluded to avoid foreground biases, an expansion over
full-sky spherical harmonic basis functions will no longer yield orthonormal
modes.
The spherical harmonic coefficients, or pseudo-$\widetilde a_{\ell m}$s,
obtained in this way will be statistically correlated between modes
$m$ {\sl and} $\ell$ in the sense that
$\langle \widetilde a_{\ell m}^{\,}\widetilde a_{\ell^\prime m^\prime}^{\ast}\rangle\neq\delta_{\ell\ell^\prime}\delta_{mm^\prime}\widetilde C_\ell$
where $\widetilde C_\ell$ is
the cut-sky, pseudo angular power spectrum defined as
\begin{equation}\label{eq:pcl}
\widetilde C_\ell = \frac{1}{2\ell+1}\sum_{m=-\ell}^{\ell} |\widetilde a_{\ell
  m}^{\,}|^2\,.
\end{equation}
The tilde ($\widetilde{\phantom{a}}$) indicates that the quantity is computed in the partial sky, filtered and beam-smoothed reference space of the data from the instrument.

\citet{master} show how the geometry of the mask applied to the data
can be used to calculate the coupling between $\widetilde a_{\ell m}^{\,}$ coefficients.
In turn, under an assumption of isotropy, this can be used
to define a linear relationship between the ensemble average of cut- and
full-sky angular power spectra:
\begin{equation}
  \label{eq:cut_sky_kernel}
  \langle \widetilde C_\ell \rangle = \sum_{\ell^\prime}K_{\ell \ell^\prime}C_{\ell^\prime}\,,
\end{equation}
where $\langle a_{\ell m}^{\,}a_{\ell^\prime m^\prime}^{\ast}\rangle =
\delta_{\ell\ell^\prime}\delta_{mm^\prime}C_\ell$ holds on the full sky and
$K_{\ell \ell^\prime}$ is a coupling kernel that can be computed from the sky
mask.  This expression can be generalized to include polarization
\citep{Challinor_2005,xfaster_planck}.

Assuming the full-sky $a_{\ell m}$ coefficients are Gaussian distributed,
then the cut-sky $\widetilde a_{\ell m}$s must also be Gaussian since they
are related by a linear transformation. We can then write the likelihood, $L$, as
\begin{equation}\label{eq:likelihood}
L( \pmb{\widetilde d}\,|\, \pmb{\theta})  = \frac{1}{\sqrt{2 \pi \, |\pmb{\widetilde C}|}} \exp
\left(-\frac{1}{2} \pmb{\widetilde d}^\dagger \cdot \pmb{\widetilde C}^{-1}
\cdot \pmb{\widetilde d}\right)\,,
\end{equation}
where $\pmb{\widetilde d}$ is a generalized data vector containing the
observed $\widetilde a_{\ell m}^{\,}$, $\pmb{\theta}$ is a vector of model parameters,  and the generalized covariance matrix
$\pmb{\widetilde C}$ is the sum of the signal and noise components of the model,
\begin{equation}
  \pmb{\widetilde C}(\theta)= \pmb{\widetilde S}(\theta)+\pmb{\widetilde N} \,,
  \label{eq:model}
\end{equation}
where $\pmb{\widetilde S}$ is the signal, which depends on the model parameters,
and $\pmb{\widetilde N}$ is the noise.  The likelihood for the data given the
parameters $\pmb{\theta}$ can be interpreted as the likelihood for the
parameters given the data $L(\pmb{\theta}\,|\,\pmb{\widetilde d})$ assuming
uniform Bayesian priors in $\pmb{\theta}$.

In principle, the exact likelihood of Equation~\ref{eq:likelihood} can be used to estimate model parameters by defining the full, non-diagonal $\ell$, $m$ structure of the covariance $\pmb{\widetilde C}$.
In practice this is not feasible because of the size of the covariance matrix and the difficulty in defining the full anisotropic structure of the noise term.

XFaster approximates the likelihood in Equation~\ref{eq:likelihood} using two simplifications.
The first is an assumption of isotropy, using Equation~\ref{eq:pcl} to assign equal variance to $m$ modes for each multipole $\ell$.
The second is to construct the model covariance using power averaged
over bins in multipoles, or ``bandpowers''. Averaging the power in
this way reduces the effect of correlations between multipoles induced by the partial coverage.
XFaster uses bandpower parameters that retain the full $\ell$, $\ell^\prime$ coupling but approximates the likelihood as diagonal in $\ell$.
This results in an unbiased estimate of the model parameters, but the effective degrees of freedom in the XFaster likelihood must be calibrated using simulations in order to obtain a robust estimate of the likelihood curvature.
This calibration, which is most important at the lowest multipoles where the effect of mode coupling is strongest, is discussed further in Section~\ref{subsec:g_ell}.

\subsection{Bandpower Model}

We parameterize the signal portion of the model (Equation~\ref{eq:model}) by introducing bandpower deviations, $q^{XY}_b$, where $b$ is a generalized index indicating the multipole range and $XY$ indicates the cross-spectrum polarization combination, \emph{i.e.},\ $TT$, $TE$, $EE$, $BB$, $EB$, $TB$.
We then construct the signal model bandpowers that the $q_b$ factors modify as described in Sections~\ref{sec:shape_spec}-\ref{sec:kernels} below.
The parameterization of the noise portion of the model is described in Section~\ref{sec:noise_model}.

\subsubsection{Shape Spectrum}
\label{sec:shape_spec}

\begin{figure}
  \includegraphics[width=\columnwidth]{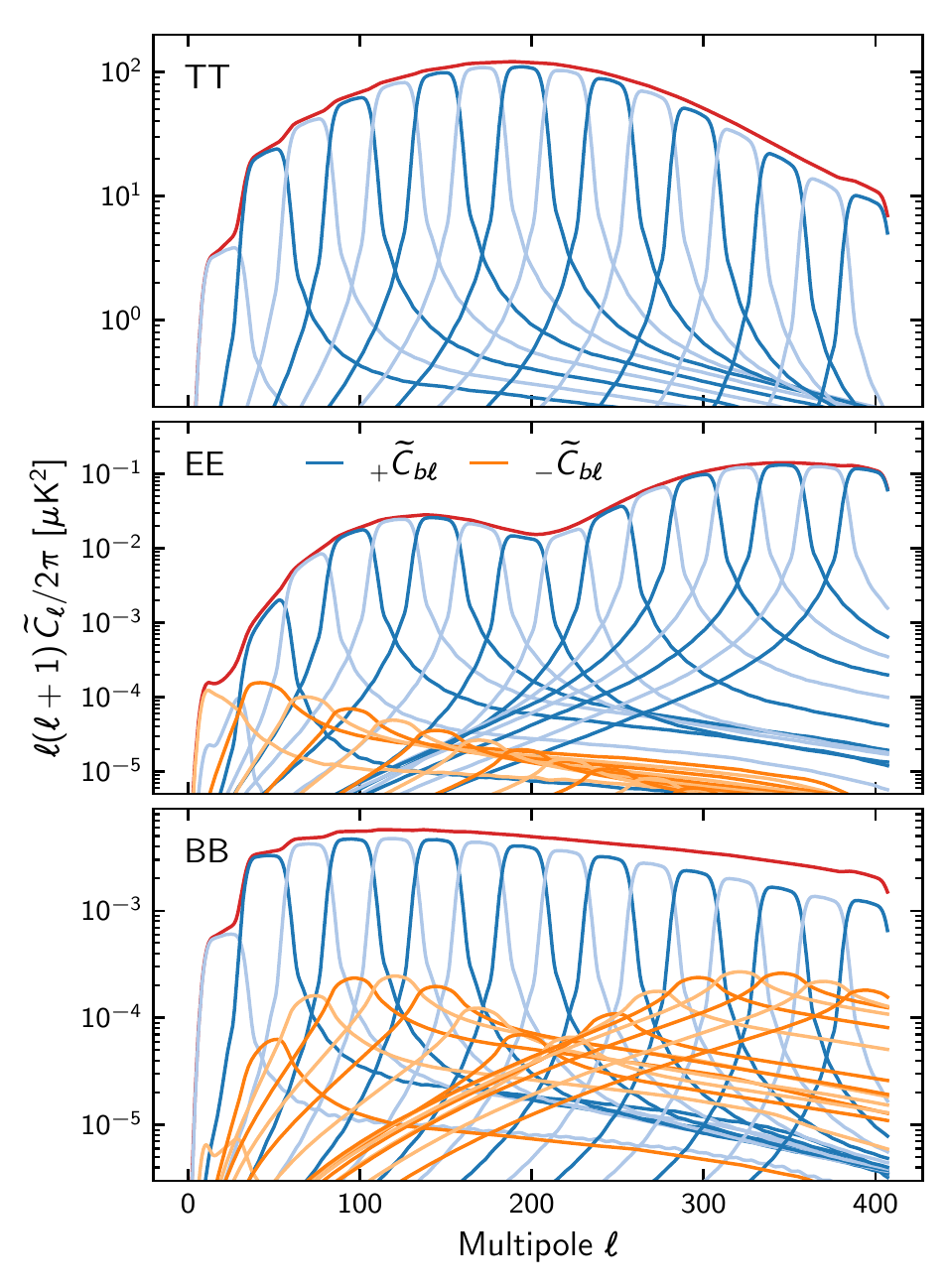}
  \caption{The CMB bandpower kernels ($\widetilde{C}_{b\ell}$) for $TT$, $EE$, and $BB$ cross-spectra, including \spider \SI{150}{\giga\hertz} masking, filtering and beam smoothing.  The binning operator $\chi_{b\ell}$ is piecewise linear with equal-sized bins of width $\Delta\ell = 25$.
    The colors alternate by bin, with the sum of the contributions from each bin (\emph{i.e.},\ the signal model $\widetilde{S}_\ell$ with $q_b = 1$) given by the red line.
    The underlying $BB$ shape spectrum is constant in $\ell (\ell + 1) C_\ell$ to have appreciable input power for determining the filter transfer function.
    The estimated data spectra have been found to be insensitive to the choice of shape spectra.
    The mode-coupling matrix terms that mix $E$ and $B$ polarizations ($_-\widetilde{C}_{b\ell}$) are also shown (\emph{orange}); these mixing terms contribute additional power in the tails for each bin, most visible in the $BB$ model.
  }
  \label{fig:cbls}
\end{figure}

The bandpower deviations $q_b$ are defined with respect to a template, full-sky angular power spectral shape $C_\ell^{(S)}$ such that any sufficiently smooth model power spectrum can be expressed as
\begin{equation}
  C^{XY}_\ell = \sum_b \chi^{XY}_{b\ell}\, q_b^{XY}\, C_\ell^{XY\,(S)}\,,
  \label{eq:signal}
\end{equation}
where $\chi^{XY}_{b\ell}$ is a binning operator that is non-zero only for the
spectrum component $XY$, and is usually assumed to be piecewise linear by
multipole range, but could be chosen to be a set of tapered, overlapping kernels.
The shape spectrum $C_\ell^{(S)}$ for the CMB is computed using the
\texttt{CAMB} package \citep{camb}.
The shape spectrum could also include other sky components in addition to the CMB, such as foregrounds.
Alternatively, the template shape can be flat, in which case the $q_b$ parameters are interpreted as the traditional bandpowers, $C^{XY}_b$.
The choice of a flat shape spectrum is appropriate for spectra that vary little within a bandpower, and is nonetheless unbiased in the mean for any spectrum shape.
However, this is a suboptimal weighting of the power if the signal
varies substantially within each bandpower. The calculated coupling between multipoles due to the
mask is also more accurate if a known template shape can be used.

\subsubsection{Signal Bandpower Kernels}
\label{sec:cbls}

The model signal component $\pmb{\widetilde S}$ for the XFaster likelihood covariance $\pmb{\widetilde C}$ can be defined using so-called bandpower kernels $\widetilde C_{b\ell}^{XY, ij}$:
\begin{equation}
  \widetilde S_\ell^{XY,ij} = \sum_b q^{XY}_b\, \widetilde C_{b \ell}^{XY,ij}
  \label{eq:pseudo_signal}
\end{equation}
We have also introduced the index combination $ij$ to indicate the cross-correlation of modes from separate maps $i$ and $j$.
This allows maps from different observations to be combined into a single estimate of a unified power spectrum.
The maps can be of different sizes, have different geometries and/or weightings, be overlapping or non-overlapping, and have different beam smoothing and transfer functions.
The coupling between modes is propagated through the estimation by the mode-coupling kernels $K_{\ell\ell^\prime}$ introduced in Equation~\ref{eq:cut_sky_kernel}.
This structure can also allow for maps of observations at different frequencies when fitting for galactic foregrounds, as described in Section~\ref{sec:harmonic_fg}.

Following the MASTER formalism, the bandpower kernels are written as:
\begin{equation}
  \widetilde{C}_{b \ell}^{XY,ij} = \sum_{\ell^\prime}\chi^{XY}_{b\ell^\prime}\, K^{ij}_{\ell \ell^\prime} F_{\ell^\prime}^{XY,ij} \left(B_{\ell^\prime}^{XY,ij}\right)^2 C_{\ell^\prime}^{XY\,(S)} \,,
  \label{eq:cbl_tt}
\end{equation}
with time-domain filter transfer function $F_\ell^{XY,ij}$, beam window function $B_\ell^{XY,ij}$, mode-coupling kernels $K_{\ell\ell^\prime}^{ij}$, and shape spectrum $C_\ell^{XY\,(S)}$.
When combined with Equation~\ref{eq:pseudo_signal}, it becomes clear that the signal model is simply Equation~\ref{eq:signal} reconstructed on the cut sky.

Equation~\ref{eq:pseudo_signal} is valid for $XY \in \{ TT,\,TE,\,TB,\,EB\}$.
However, the cut-sky mask results in the mixing of $E$- and $B$-modes, which must be accounted for in their spectral models.
The remaining spectral components are:
\begin{equation}
  \widetilde S_\ell^{XX,ij} = \sum_b q^{XX}_b\, _+\widetilde C_{b \ell}^{XX,ij}+\sum_b q^{YY}_b\, _-\widetilde C_{b \ell}^{YY,ij}\,,
  \label{eq:pseudo_signal_eb}
\end{equation}
where now $XX$ and $YY$ are the combinations $EE$ and $BB$, and the $_-\widetilde{C}_{b\ell}$ terms mix $BB$ power into the $EE$ signal model and vice-versa.

Equation~\ref{eq:cbl_tt} is valid as written for $XY = TT$, but must be modified to construct the remaining bandpower kernels.
The $EE$ kernel terms are:
\begin{equation}
  \begin{aligned}
    {}_{+}\widetilde{C}_{b \ell}^{EE,ij} &= \sum_{\ell^\prime} \chi^{EE}_{b\ell^\prime}\, {}_{+} K^{ij}_{\ell \ell^\prime} F_{\ell^\prime}^{EE,ij} \left(B_{\ell^\prime}^{EE,ij}\right)^2 C_{\ell^\prime}^{EE\,(S)}\,, \\
    {}_{-}\widetilde{C}_{b \ell}^{EE,ij} &= \sum_{\ell^\prime} \chi^{BB}_{b\ell^\prime}\, {}_{-} K^{ij}_{\ell \ell^\prime} F_{\ell^\prime}^{EE,ij} \left(B_{\ell^\prime}^{EE,ij}\right)^2 C_{\ell^\prime}^{EE\,(S)}\,, \\
  \label{eq:cbl_ee}
  \end{aligned}
\end{equation}
and similarly for $BB$, where we now introduce the polarization mode-coupling kernels ${}_\pm K^{ij}_{\ell \ell^\prime}$.  In particular, ${}_- K^{ij}_{\ell \ell^\prime}$ accounts for the $E-B$ mixing terms.
The $EB$ cross bandpower kernels are:
\begin{equation}
  \widetilde{C}_{b \ell}^{EB,ij} = \sum_{\ell^\prime} \chi^{EB}_{b\ell^\prime}\,\left({}_+ K^{ij}_{\ell \ell^\prime}-{}_- K^{ij}_{\ell \ell^\prime}\right) F_{\ell^\prime}^{EB,ij} \left(B_{\ell^\prime}^{EB,ij}\right)^2 C_{\ell^\prime}^{EB\,(S)} \,.
\label{eq:cbl_eb}
\end{equation}
Finally for $XY \in \{TE,\,TB\}$ we have:
\begin{equation}
  \widetilde{C}_{b \ell}^{XY,ij}=\sum_{\ell^\prime} \chi^{XY}_{b\ell^\prime}\,{}_\times K^{ij}_{\ell \ell^\prime} F_{\ell^\prime}^{XY,ij} \left(B_{\ell^\prime}^{XY,ij}\right)^2 C_{\ell^\prime}^{XY\,(S)} \,,
  \label{eq:cbl_te}
\end{equation}
where the mode-coupling kernel ${}_\times K^{ij}_{\ell \ell^\prime}$ describes the coupling between temperature and polarization.

The $\widetilde C_{b \ell}$s for $TT$, $EE$, and $BB$ CMB shape spectra for the \spider \SI{150}{\giga\hertz} cross-spectrum are shown in Figure \ref{fig:cbls}.
The computation of the components of the bandpower kernels, and their values for \spider, are given in Sections~\ref{sec:beams}-\ref{sec:kernels}.

\subsubsection{Beam Window Functions}
\label{sec:beams}

\begin{figure}
  \centering
  \includegraphics[width=\columnwidth]{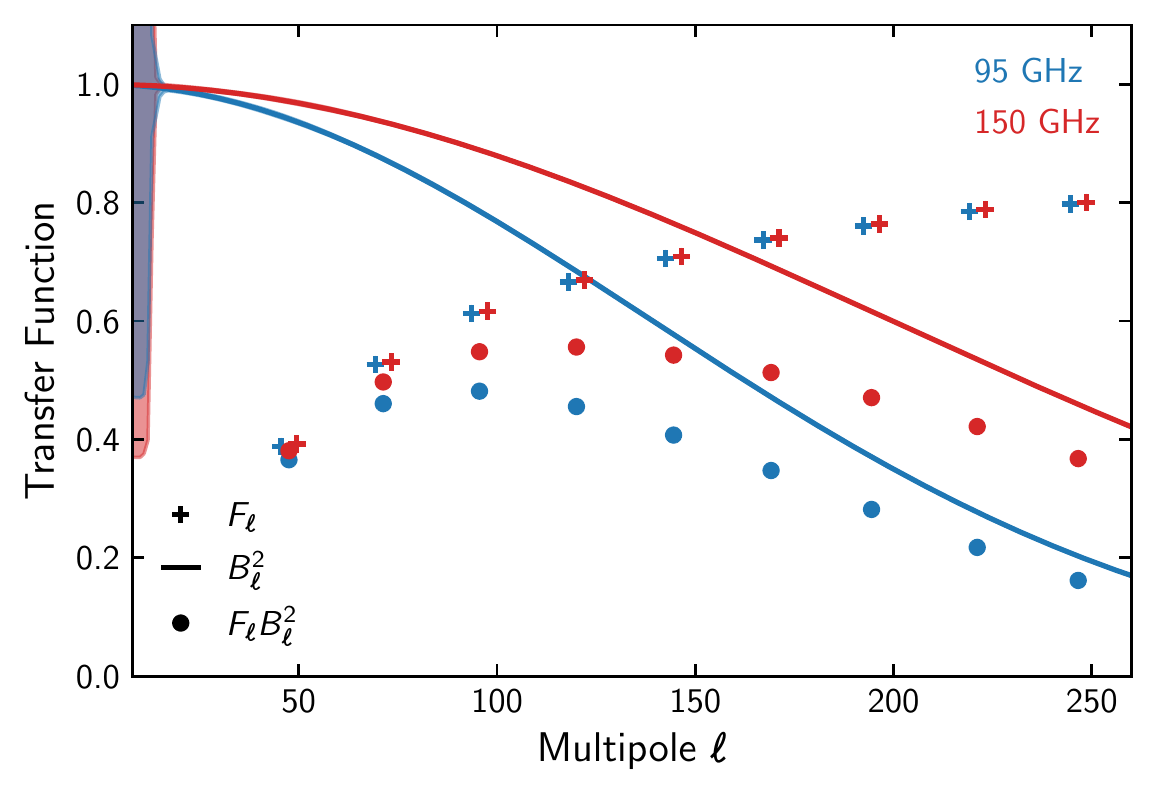}
  \caption{\spider's filter transfer function ($F_\ell$), beam window function ($B_\ell^2$), and total
    transfer function for \SIlist{95;150}{\giga\hertz}.
    Beam error envelopes at each frequency are also shown, most evident at low multipoles.
    The envelopes represent the $1\sigma$ statistical uncertainty in the beam transfer function as determined by the scatter
    in per-detector beam estimates.
    Quantities shown are the average of the $EE$ and $BB$ transfer functions, which are similar but are not assumed
    to be identical.}
  \label{fig:transfers_beams}
\end{figure}

The beam window functions $B_\ell$ are an input to the XFaster algorithm.
One window function is required per map, and the estimated error on the beam can also be input to the pipeline.
The error may be marginalized over in computing the cosmological parameter likelihoods to account for these uncertainties.

The beam terms in the bandpower kernels are constructed as the product of the individual beam windows for each of the two maps indexed by $i$ and $j$:
\begin{equation}
  \left(B_\ell^{XY,ij}\right)^2 = B_\ell^{XY,i} \cdot B_\ell^{XY,j}\,,
  \label{eq:beams}
\end{equation}
and the beam error terms are included by adding derivatives of the model with respect to each beam window to the signal covariance.

For \spider, the beam window functions are computed by cross-correlating \spider data maps at \SIlist{95;150}{\giga\hertz} with \planck maps at \SIlist{100;143}{\giga\hertz}, respectively\footnote{Throughout this paper we use release~3.01 of the \planck HFI maps \citep{planck18_hfi}}.
The beam is modeled as a Gaussian function, with an approximate FWHM of 41\,arcmin at \SI{95}{\giga\hertz} and 29\,arcmin at \SI{150}{\giga\hertz}.
The errors on the average beams are determined from the distribution of estimated detector beams at that frequency.
This produces a $1\sigma$ Gaussian error envelope as a function of $\ell$.
The error envelope acts as a Gaussian prior on the beam shape when computing parameter likelihoods.
The \texttt{HEALPix} $N_\mathrm{side}=512$ pixel window function is multiplied by the instrument beam window function to account for smoothing from pixelization in the process of producing pixelized maps from discretely-sampled time-ordered data.
The beam window functions and $1\sigma$ statistical errors for \spider are shown in Figure \ref{fig:transfers_beams}.

\subsubsection{Filter Transfer Functions}
\label{sec:filter_transfer_function}

In practice, observations of the sky are binned into maps from time-ordered data, which must be filtered to remove systematics like scan-synchronous noise or noisy frequencies.
This filtering suppresses signal modes at certain angular scales, and the resulting bias must be computed empirically by comparing an input model spectrum to the spectra of an ensemble of simulations which have been filtered identically to the on-sky data.
As in the MASTER formalism, we approximate the filter transfer function $F_\ell$ as a spherically symmetric function of only $\ell$-modes.
We also assume that the transfer function is independent of the input signal spectrum used to compute it.
We have verified that this is a good assumption for CMB spectra, and in particular that the $BB$ CMB spectra are insensitive to reasonable changes to input spectra.
However, using CMB-like input spectra to compute transfer functions results in a bias at low multipoles for maps with a significant dust-like foreground component.
Further efforts to account for this bias in the estimation of dust bandpowers will be presented in future work.

The filter transfer functions $F_\ell$ are computed in the same way as bandpowers, described in Section~\ref{sec:like_max}, but substituting the average of signal-only simulations for the observed data, and setting the $F_\ell$ term in the signal model to 1.
The remaining non-unitarity of the $q_b$ values is the binned transfer function.
Transfer functions are computed for $TT$, $EE$, and $BB$ spectra, independently for each map, since the filtering may differ significantly between maps.
$TE$, $EB$, and $TB$ transfer functions are approximated as the geometric mean of their component transfer functions, \emph{e.g.}, $F^{TE}_\ell=\sqrt{{F^{TT}_\ell F^{EE}_\ell}}$.

When constructing the signal model using the binned transfer function, we expand $F_b$ to the full $\ell$ range using a constant value in each bin.
The transfer function term for each cross-spectrum in the signal model is then the geometric mean of the transfer functions for each of the two maps indexed by $i$ and $j$:
\begin{equation}
  F_\ell^{XY,ij} = \sqrt{F_\ell^{XY,i} \cdot F_\ell^{XY,j}}\,.
\end{equation}
The \spider $EE$ and $BB$ transfer functions are shown in Figure \ref{fig:transfers_beams}.

\subsubsection{Mode-Coupling Kernels}
\label{sec:kernels}

\begin{figure}
  \centering
  \includegraphics[width=\columnwidth]{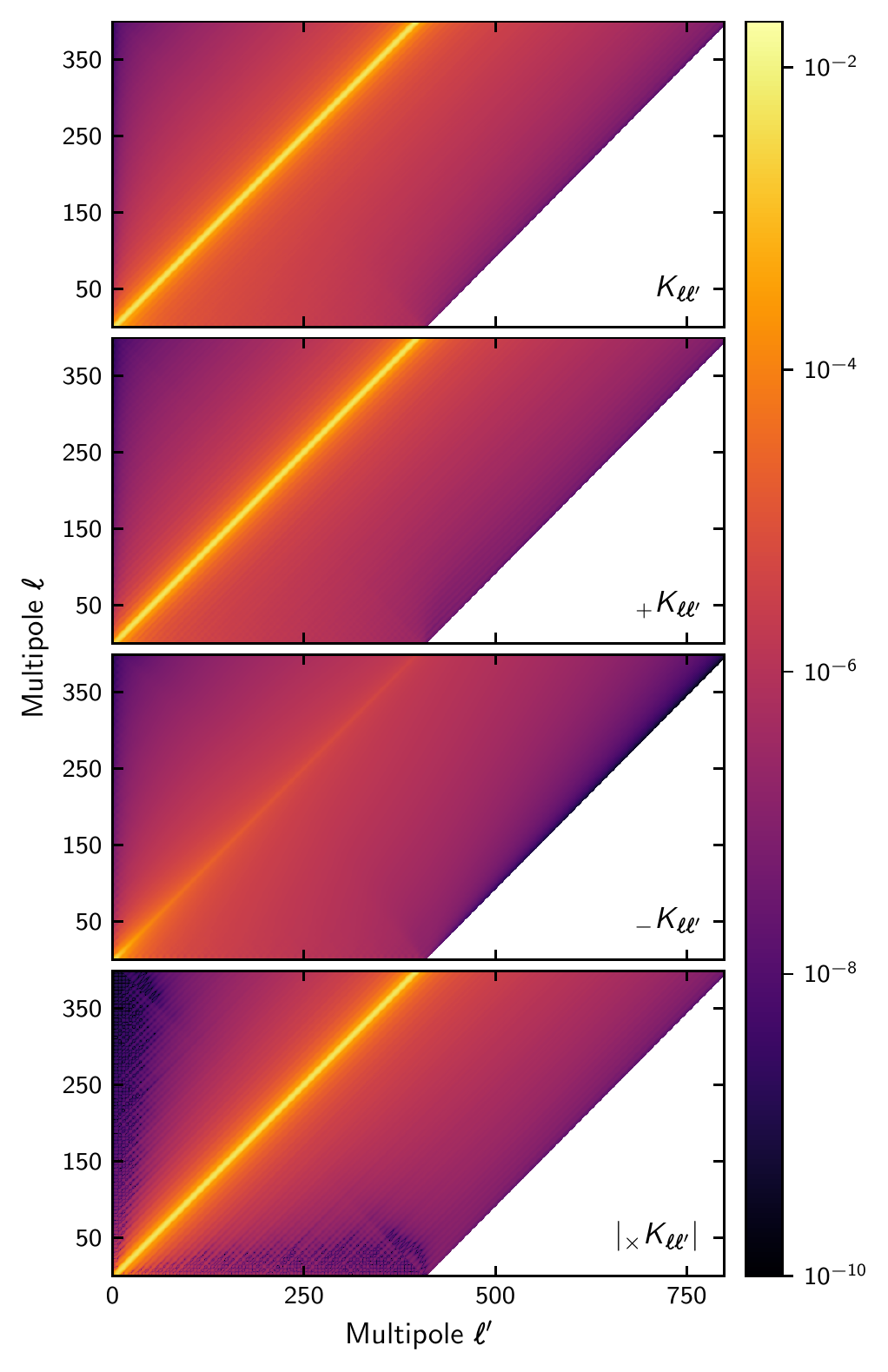}
  \caption{The mode-coupling kernels, $K_{\ell\ell^\prime}$, for the \spider mask.
    Each subplot corresponds a component in Equation~\ref{eq:kernels}.
    For plotting purposes, negative elements are set to their absolute value for ${}_\times K_{\ell\ell^\prime}$, resulting in the low-$\ell$ structure seen.}
  \label{fig:kernels}
\end{figure}

Multiple mode-coupling kernels are required in the case of polarization-sensitive observations or maps with different masks.
The weighting applied to masked maps can also be different for each Stokes parameter $I$, $Q$, and $U$.
The kernels are computed from the cross-correlation power spectrum of the masks on the full sky, $\mathcal{W}_L$, and we compute separate kernels for each of the polarization combinations, and for each unique pair of maps $i$ and $j$. Following \cite{Challinor_2005}, the extended set of polarization-sensitive kernels are:
\begin{equation}
\begin{aligned}
K^{ij}_{\ell\ell^\prime} &=\!\! \frac{2 \ell^\prime + 1}{4 \pi} \sum_L (2 L + 1) \,\mathcal{W}^{TT,ij}_L
\!\!\begin{pmatrix}
\ell & \ell^\prime & L \\
0 & 0 & 0
\end{pmatrix}^2 \,,\\
_\pm K^{ij}_{\ell\ell^\prime} &=\!\!  \frac{2 \ell^\prime + 1}{8 \pi} \sum_L(2 L + 1) \,\mathcal{W}^{PP,ij}_L
\!\!\begin{pmatrix}
\ell & \ell^\prime & L \\
2 & -2 & 0
\end{pmatrix}^2 \!\!\left[1 \pm (-1)^{\ell + \ell^\prime + L}\right] \,,\\
_\times K^{ij}_{\ell\ell^\prime} &=\!\!  \frac{2 \ell^\prime + 1}{4 \pi} \sum_L (2 L + 1) \,\mathcal{W}^{TP,ij}_L
\!\!\begin{pmatrix}
\ell & \ell^\prime & L \\
2 & -2 & 0
\end{pmatrix} \begin{pmatrix}
\ell & \ell^\prime & L \\
0 & 0 & 0
\end{pmatrix}\,,
\end{aligned}
\label{eq:kernels}
\end{equation}
where the terms in parentheses are the Wigner 3-j symbols, $T$ is the Stokes $I$ mask, and $P$ is the Stokes $Q$/$U$ mask which need not be identical.
The $\pm$ kernels are used to compute $EE$, $BB$ and $EB$ power spectrum terms, with the $-$ term in particular used to account for mixing between $E$ and $B$ due to the mask.
The $\times$ kernel is used to compute the $TE$ and $TB$ spectra. The kernels for the \spider mask are shown in Figure \ref{fig:kernels}.

\subsubsection{Residual Noise Calibration}
\label{sec:noise_model}
The XFaster likelihood approximation also enables an estimation of noise calibration parameters.
Uncorrelated noise enters the covariance matrix as a diagonal term.
To account for inaccuracies in our noise simulations, we fit a scalar parameter $n_b^i$ per bin as a correction to the noise model:
\begin{equation}
\label{eq:noise_component}
    \widetilde{N}^{XY,ij}_\ell= \delta^{ij} \sum_b \, \chi^{XY}_{b\ell} \left(1 + n_{b}^i\right) \langle\,\widetilde{N}_{\ell}^{i}\,\rangle\,,
\end{equation}
where $\langle\,\widetilde{N}_{\ell}^{i}\,\rangle$ is the mean spectrum of an ensemble of noise simulations.
Noise is treated differently from signal, in that we model it directly in the cut-sky power spectrum with no polarization coupling.
In principle, noise is coupled across polarizations, but this is difficult to account for analytically in the cut-sky spectra.

Clearly, if the binning structure of the $q_b$s and the $n_b$s are too similar and/or if the template biases $\langle\,\widetilde{N}_{\ell}^{i}\,\rangle$ have a similar $\ell$ dependence to the shape templates $C_\ell^{XY}$, this extension will introduce significant degeneracies in an auto-spectrum analysis.
The degeneracies are broken by including multiple cross-spectra, where noise biases do not contribute.
This has been found for \spider to sufficiently decouple the signal and noise parameters, though increasing the bin width for the $n_b$ parameters in comparison to that for $q_b$ would further address this potential issue.
The piecewise linear model for the noise calibration means the $n_b$ ``noise residual'' parameters can be estimated jointly with the signal $q_b$s and can be marginalized over using the full Fisher matrix, once the estimator has converged to the maximum likelihood solution.

\subsection{Likelihood Computation}

The components introduced above are used to construct the signal and noise covariances in Equation~\ref{eq:likelihood}.
In the XFaster approximation the covariances are block-diagonal by $\ell$.
The sub-blocks are built from the cross-spectra of \textit{N} maps and polarizations with each sub-block as follows:
\begin{equation}
\widetilde{\pmb{C}}_{\ell}^{ij} =
\left[\begin{array}{ccc}{
\widetilde{\mathrm{C}}_{\ell}^{T T}} &
\widetilde{\mathrm{C}}_{\ell}^{T E} &
\widetilde{\mathrm{C}}_{\ell}^{T B} \\
- &
\widetilde{\mathrm{C}}_{\ell}^{E E} &
\widetilde{\mathrm{C}}_{\ell}^{E B} \\
- &
- &
\widetilde{\mathrm{C}}_{\ell}^{B B}
\end{array}\right]_{ij}\,,
\end{equation}
where $i$ and $j$ index over the $N$ independent maps.
Formulated in this way, the matrix $\widetilde{\pmb{C}}_\ell$ is the \textit{covariance} of the $\widetilde a^{X,i}_{\ell m}$s that make up the generalized, observed data vector $\widetilde{\pmb{d}}$ in Equation~\ref{eq:likelihood}.
The block-diagonal form of the covariance means the data vector can be pre-compressed into spectra:
\begin{equation}
\widehat{C}_\ell^{XY,ij} = \frac{1}{2\ell+1}\sum_m \widehat{a}^{X,i}_{\ell m} \widehat{a}^{\ast \, Y,j}_{\ell m} \,.
\end{equation}
Here and elsewhere, we use the hat symbol ($\widehat{\phantom{a}}$) to distinguish matrices of \textit{data} pseudo-spectra from general matrices of pseudo-spectra.  Then the log-likelihood, up to an overall constant, can be written as
\begin{equation}
 \mathcal{L}\equiv \ln \,L = -\frac{1}{2} \sum_{\ell k} (2 \ell + 1) g_\ell^k \left[\widetilde{\pmb{C}}_\ell^{-1} \cdot\widehat{\pmb{C}}_\ell + \ln\,\widetilde{\pmb{C}}_\ell\right]_{kk} \,,
  \label{eq:logL}
\end{equation}
where $k$ indexes over polarization and independent maps in the sub-blocks (see Equation 2 in \cite{bond2000}).
The factor $(2\ell + 1)$ appears as a degree of freedom count due to the block-diagonal form. The coefficient $g_\ell^k$ accounts for the {\sl effective} number of modes from each map that contribute to the final trace for each multipole.

\subsubsection{Likelihood Maximization}
\label{sec:like_max}
The XFaster likelihood (Equation~\ref{eq:logL}) can be maximized using
an iterative quadratic estimator \citep{bjk} to find the maximum likelihood estimates for parameters $\theta$ which include all $q_b$s and $n_b$s that are allowed to vary freely.
The method uses the Fisher matrix to approximate the curvature of the likelihood at each iteration. The Fisher matrix is given by
\begin{equation}
  \mathcal{F}_{bb^\prime} = \frac{1}{2}\sum_{\ell k} (2\ell + 1)\, g_\ell^k\left[\frac{\partial \widetilde{\pmb{C}}_\ell}{\partial \theta_b} \cdot \widetilde{\pmb{C}}^{-1}_\ell \cdot \frac{\partial \widetilde{\pmb{C}}_\ell}{\partial \theta_{b^\prime}} \cdot \widetilde{\pmb{C}}^{-1}_\ell \right]_{kk}\,,
  \label{eq:fisher}
\end{equation}
where we have used the same notation convention as in Equation~\ref{eq:logL} and the index $b$ now runs across all parameters in the set of $q_b$s and $n_b$s.
In practice, starting from an initial guess with $\theta_b=1$, we calculate an updated solution at each iteration using
\begin{equation}
\label{eq:q_bestimator}
\theta_b = \frac{1}{2}\sum_{b^\prime \ell k}\mathcal{F}_{bb^\prime}^{-1}(2\ell + 1)\,g_\ell^k \left[\widetilde{\pmb{C}}^{-1}_\ell \cdot \frac{\partial \widetilde{\pmb{C}}_\ell}{\partial \theta_{b^\prime}} \cdot \widetilde{\pmb{C}}^{-1}_\ell \cdot \left(\widehat{\pmb{C}}_\ell-\widetilde{\pmb{N}}_\ell \right)  \right]_{kk}\,.
\end{equation}
This is equivalent to a Newton-Raphson minimization method.
Iterations can be terminated when a convergence criterion is satisfied.
We terminate when the maximum of the absolute fractional change in $\theta_b$ is below some threshold, typically $10^{-3}$.  Note that the mode-counting factor $g_\ell$ enters into the elements of the Fisher matrix (and therefore, the resulting uncertainties on the parameters), but is effectively divided out in Equation~\ref{eq:q_bestimator}, so mis-calibration may result in biased uncertainties but {\sl not} biased parameter estimates.

In practice, the model matrix is not positive definite because the
bandpower deviations are allowed to be negative. This is not strictly
a problem since the likelihood is not evaluated during the
optimization, but the problem becomes ill-conditioned if the steps
approach the threshold where the covariance becomes singular. This
leads to spurious values for the gradient contribution, driven by
numerical errors, and can slow down or prevent convergence. The problem
can be solved by ensuring the matrices involved are better
conditioned.  This is achieved by adding a conditioner to the
diagonals of the covariances and is equivalent to adding a numerical
floor to the eigenvalues of the matrices.  In our implementation, the
conditioning level is introduced if the bandpower deviations are
failing to converge and then adjusted automatically to the minimum
level required for convergence.  The conditioning is then dropped when
estimating the final Fisher matrix, at the likelihood maximum, since the
matrices are always invertible if the likelihood point is
well-defined.\footnote{See \cite{Gjerlow} for a similar approach
  using a conditioning prior and an additional step regularizer.}

The conditioning step is also a
useful indicator of systematics.  If the presence of the conditioner
leads to  significant fractional changes to the value of the likelihood at the
maximum, this can indicate that inconsistencies in the observed data
are driving the
spurious contributions to the gradient term in Equation
\ref{eq:q_bestimator}. This is a particularly useful check when
analyzing multiple maps.

\subsubsection{Mode-Counting Factor, $g_\ell$}
\label{subsec:g_ell}

\begin{figure}
  \centering
  \includegraphics[width=\columnwidth]{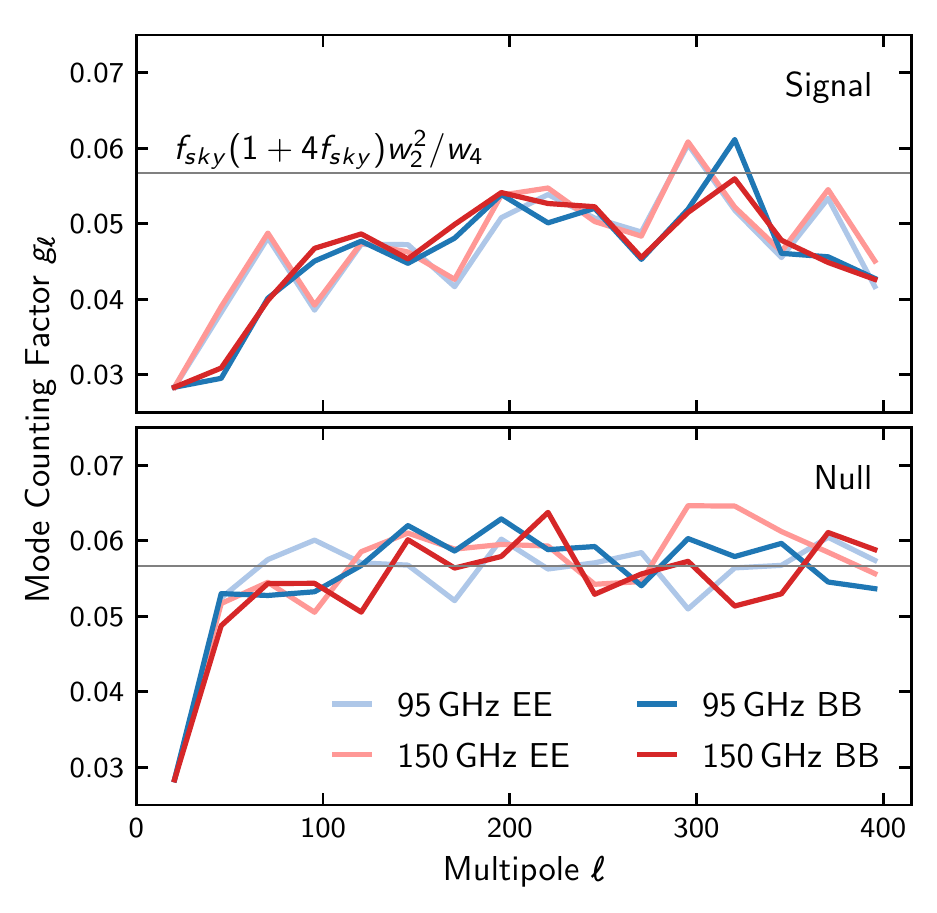}
  \caption{The $g_\ell$ effective mode-counting factor as estimated for \spider.
    The predominant effect is a constant approximate 6\% retention of the full sky power after masking (Equation~\ref{eq:gm}).
    An additional $\ell$-dependent reduction results from ``missing'' contributions to the covariance due to approximations in the construction of the likelihood, as described in Section~\ref{subsec:g_ell}.
    This component is empirically calibrated with simulations.
  }
  \label{fig:gcorr}
\end{figure}

The mode-counting factor $g_\ell$, introduced in Equation~\ref{eq:logL}, accounts for the change to the effective degrees of freedom in the likelihood induced by both map weighting {\sl and} the ``missing'' contributions to the covariance in the block-diagonal likelihood approximation.
It is important to accurately calibrate this factor when using the XFaster method, in order to produce unbiased uncertainties on the fitted parameters in the signal model.

The mode loss is most pronounced at low multipoles, approaching the overall scales of each map, where $\ell$-to-$\ell^\prime$ couplings are significant.
In dealing with this bias, one option is to limit the range in $\ell$ covered by the $q_b$ parameters, but we have found the addition of a $g_\ell$ factor significantly extends the useful range of multipoles where bandpowers can be estimated accurately.

To calculate $g_\ell$, we first estimate an overall $\ell$-independent starting amplitude based on the ratio of mask moments introduced in \cite{master}:
\begin{equation}\label{eq:gm}
  g = f_{sky}\frac{w_2^2}{w_4}\,,
\end{equation}
where $w_n$ is the $n^{th}$ moment of mask $w_p$ with
\begin{equation}
w_n=\frac{1}{f_{sky}}\sum_p w^n_p\,,
\end{equation}
where $p$ is the pixel index for the mask and $f_{sky}$ is the fraction of the sky covered by the mask.
For past applications, such as \cite{Netterfield2002} and \cite{Montroy:2005yx}, this was found to be a sufficient approximation to accurately model the variance of the final bandpowers.
We have found that the application to \spider requires a more accurate counting of modes, likely due to the presence of high signal-to-noise polarization modes at large scales for which the coupling structure is most complicated.
The mode loss induced by the coupling can, in principle, be computed analytically to higher order in map moments (see \cite{Challinor_2005}) but the calculation is not straightforward and relies on a number of simplifying assumptions.
In practice, we find a second-order correction to the overall amplitude $g$ by a factor of $(1+4f_{sky})$, combined with an empirical Monte Carlo estimate of the $\ell$-dependence, is required for the correct calibration of the Fisher matrix when compared to end-to-end simulations.

Since the mode-coupling factor is partially degenerate with the filter transfer function $F_\ell$, we estimate the final $g$-correction iteratively by computing bandpowers for 1000 signal-only simulations that have been filtered in the same way as the data, and compare the scatter of the ensemble of bandpowers to the diagonal of the Fisher matrix (\emph{i.e.}, XFaster's estimate of the error bar).
We use the ratio as an estimate of the $g$-correction, feed it back into the estimate of the ensemble of bandpowers and repeat the process until the correction converges.

Figure \ref{fig:gcorr} shows the amplitude of the total $g_\ell$ factor per bin, using both signal simulations and null (noise-dominated) simulations.  The latter are discussed in more detail in Section~\ref{sec:nulls}.  We note here that in the noise-dominated regime, the analytical equation for $g_\ell$ with the constant second-order correction appears to be a good approximation on average at all but the largest angular scales, while the signal-dominated regime shows more structure as a function of $\ell$.

\subsection{Prior Constraints}
\label{sec:priors}

A likelihood-based estimator such as the quadratic estimator of Equation~\ref{eq:q_bestimator} is easily modified to included prior constraints on any of the parameters $\theta_b$.
Priors can help improve convergence if poorly constrained parameter directions are included.
They are also very useful to include self-consistent marginalization over prior constraints on parameters to obtain a final Fisher matrix that contains a full propagation and accumulation of errors.
This is important for nuisance parameters such as noise residual calibrations or foreground parameters (see Section~\ref{sec:extensions}).

Including Gaussian priors, which suffices for most practical applications, is particularly simple. At each iteration of the quadratic estimator the Fisher matrix and estimate values can be modified as
\begin{align}
{\mathcal F}_{bb^\prime} &\rightarrow {\mathcal F}_{bb^\prime} + \delta_{bb^\prime}\frac{1}{\sigma_b^{2}}\,,\\
\theta_{b} &\rightarrow \theta_{b} + \frac{\mu_b}{\sigma_b^2}\,,
\end{align}
where $\mu_b$ and $\sigma^2_b$ are the Gaussian means and variances for each of the priors.
In the limit of a tight prior $\sigma_b^2 \rightarrow 0$, this method thus recovers $\theta_b \rightarrow \mu_b$ upon convergence.
Final marginalization over any subset of parameters can be achieved by excluding the rows and columns for those parameters in the final Fisher matrix before inversion to obtain the estimated covariances.

\subsection{Bandpower Window Functions}
\label{sec:windows}

XFaster is a method best suited for surveys where reduced sky coverage requires the use of a compression to bandpowers in order to reduce the effect of mode correlations.
The final estimated quantities in this case are a set of bandpowers $C_b$ or bandpower deviations $q_b$ and their associated Fisher matrices.
To compare these to any proposed model for the full-sky angular power spectrum, $C_\ell$, one needs to calculate model $C_b$ or $q_b$ values.
Bandpower window functions are needed for this step \citep{Bond_Leshouches,Knox_window_functions}.

Bandpower window functions are linear operators that transform the full-sky spectrum into the estimated quantity.
The window functions depend on the effective filtering induced by the observation strategy, the correlations induced by any sky cut, {\sl and} the definition of the estimator.
Different, unbiased estimators acting on the same set of observations,
for example, will produce estimates that are, in general, different and
will only agree in the ensemble limit.

Given a model spectrum $C_\ell$, the window functions are the weighting
operators appearing in the logarithmic averaging of the power into generalized
bandpowers.  For bandpower parameters $\theta_b$ that correspond to an
underlying spectrum $C_\ell$ in the ensemble limit, this generalizes to
\begin{equation}
  \langle \theta^{XY}_b\rangle = \sum_\ell \mathcal{N}^{\,}_\ell W^{XY(\theta)}_{b\ell} C^{XY}_\ell \,,
  \label{eq:wbl}
\end{equation}
where the weighting function $\mathcal{N}_\ell = (2\ell + 1) / 4\pi$, and the
window functions are normalized as $\sum_\ell \mathcal{N}^{\,}_\ell
W^{XY(\theta)}_{b\ell} C^{XY(M)}_\ell = 1$, where $C^{XY(M)}_\ell$ is a set of
model spectra.  When the model spectrum is that which is used to construct the
signal covariance (\emph{i.e.}, $C^{(M)}_\ell = \sum_b \chi_{b\ell}
C^{(S)}_\ell$), Equation~\ref{eq:wbl} results in bandpower deviations $q_b$;
when the model spectrum is flat (\emph{i.e.}, $C^{(M)}_\ell = 1$), the result is
a set of $C_b$ bandpowers; and when the model spectrum is scale-invariant
(\emph{i.e.}, $\ell (\ell + 1) C^{(M)}_\ell / 2\pi = 1$), the result is a set of
bandpowers with a scale-invariant weighting over $\ell$.

Comparing Equation~\ref{eq:wbl} to the XFaster estimator of
Equation~\ref{eq:q_bestimator} in the ensemble limit (where
$\langle\widetilde{C}_\ell - \widetilde{N}_\ell\rangle \rightarrow
\widetilde{S}_\ell$ and $\langle q_b \rangle \rightarrow 1$), we find that the
window functions for the bandpower deviations $q_b$ are given by
\begin{equation}
  W^{XY(q)}_{b\ell} = \frac{2\pi}{2\ell + 1}\sum_{b^\prime \ell^\prime k} \mathcal{F}_{bb^\prime}^{-1} (2\ell^\prime + 1)\,g_{\ell^\prime}^k
  \!\!\left[\widetilde{\pmb{C}}^{-1}_{\ell^\prime} \cdot \frac{\partial \widetilde{\pmb{C}}_{\ell^\prime}}{\partial \theta^{\,}_{b^\prime}}
    \cdot \widetilde{\pmb{C}}^{-1}_{\ell^\prime} \cdot \widetilde{\pmb{M}}^{XY}_{\ell^\prime \ell} \right]_{kk}\,,
  \label{eq:wbl_qb}
\end{equation}
where all quantities are evaluated at their maximum likelihood values.
$\widetilde{\pmb{M}}_{\ell^\prime \ell}$ is a matrix of derivatives of
the signal model $\widetilde{\pmb{S}}_\ell$ (Equation~\ref{eq:pseudo_signal}) with
respect to the model spectrum $C^{(M)}_\ell$.  Explicitly, each of the spectrum
component blocks are:
\begin{equation}
  \begin{aligned}
    \widetilde{M}^{TT,ij}_{\ell^\prime \ell} &= \delta^{TT} K^{ij}_{\ell^\prime \ell} F^{TT,ij}_{\ell} \left(B^{TT,ij}_{\ell}\right)^2\,,\\
    \widetilde{M}^{TE,ij}_{\ell^\prime \ell} &= \delta^{TE} {}_{\times}K^{ij}_{\ell^\prime \ell} F^{TE,ij}_{\ell} \left(B^{TE,ij}_{\ell}\right)^2\,,\\
    \widetilde{M}^{TB,ij}_{\ell^\prime \ell} &= \delta^{TB} {}_{\times}K^{ij}_{\ell^\prime \ell} F^{TB,ij}_{\ell} \left(B^{TB,ij}_{\ell}\right)^2\,,\\
    \widetilde{M}^{EE,ij}_{\ell^\prime \ell} &= \left(\delta^{EE} {}_{+}K^{ij}_{\ell^\prime \ell} + \delta^{BB} {}_{-}K^{ij}_{\ell^\prime \ell}\right) F^{EE,ij}_{\ell} \left(B^{EE,ij}_{\ell}\right)^2\,,\\
    \widetilde{M}^{BB,ij}_{\ell^\prime \ell} &= \left(\delta^{BB} {}_{+}K^{ij}_{\ell^\prime \ell} + \delta^{EE} {}_{-}K^{ij}_{\ell^\prime \ell}\right) F^{BB,ij}_{\ell} \left(B^{BB,ij}_{\ell}\right)^2\,,\\
    \widetilde{M}^{EB,ij}_{\ell^\prime \ell} &= \delta^{EB} \left({}_{+}K^{ij}_{\ell^\prime \ell} - {}_{-}K^{ij}_{\ell^\prime \ell}\right) F^{EB,ij}_{\ell} \left(B^{EB,ij}_{\ell}\right)^2\,,\\
  \end{aligned}
\end{equation}
where $\delta^{XY}$ is nonzero for only the $XY$ spectrum components of the matrix.

To compute bandpowers $C_b$ from bandpower deviations $q_b$ and their associated
window functions, we construct the underlying spectrum $C_\ell$ in terms of $q_b$ using
Equation~\ref{eq:signal}, and normalize Equation~\ref{eq:wbl} with respect to a
flat model, so that
\begin{equation}
  C^{XY}_b = \sum_B q^{XY}_B \frac{\sum_\ell \mathcal{N}^{\,}_\ell W^{XY(q)}_{b\ell} \chi^{XY}_{B \ell} C^{XY(S)}_\ell} {\sum_\ell \mathcal{N}^{\,}_\ell W^{XY(q)}_{b\ell} }
  \equiv \sum_B q^{XY}_B \frac{\partial \langle C^{XY}_b \rangle}{\partial q^{XY}_B}\,,
  \label{eq:cb_from_qb}
\end{equation}
where the second equality defines the block-diagonal matrix for ``rotating''
deviations $q_b$ into bandpowers $C_b$ for each spectrum component.  The
bandpower covariance is then
\begin{equation}
  \pmb{C}_{bb^\prime} = \sum_{BB^\prime} \mathcal{F}^{-1}_{BB^\prime} \frac{\partial \langle C_b \rangle}{\partial q_B}
  \frac{\partial \langle C_{b^\prime} \rangle}{\partial q_{B^\prime}}\,,
  \label{eq:cov_from_fisher}
\end{equation}
where $\mathcal{F}_{BB^\prime}$ is the Fisher information matrix of
Equation~\ref{eq:fisher}.  Finally, we obtain bandpower window functions as
\begin{equation}
  W^{XY(C)}_{b\ell} = \sum_B W^{XY(q)}_{B\ell} \frac{\partial \langle C^{XY}_b \rangle}{\partial q^{XY}_B}\,,
  \label{eq:wbl_cb}
\end{equation}
with normalization condition $\sum_\ell \mathcal{N}^{\,}_\ell W^{XY(C)}_{b\ell} = 1$.
Equations~\ref{eq:wbl} and~\ref{eq:wbl_cb} can be used to evaluate a
likelihood point for a spectrum $C_\ell$ from a set of cosmological parameters.

\subsection{Parameter Likelihood}
\label{sec:ParamLikelihood}

The definition of an approximation for the data likelihood (Equation~\ref{eq:logL}) introduces the possibility of circumventing the power spectrum and associated bandpowers altogether.
Since the space of models can be scanned directly as a function of cosmological parameters $\pmb{a}$, we can define a likelihood for $\pmb{a}$, given the data $\pmb{\widetilde d}$, using the Bayesian chain formalism
\begin{equation}
\label{eq:DirectLikelihood}
L(\pmb{a}|\pmb{\widetilde d}) \sim  P(\pmb{a})\, L(\pmb{\widetilde d}|\pmb{a})\,,
\end{equation}
where $P(\pmb{a})$ is the prior in the cosmological parameters.

For the analysis of \spider data, the only cosmological parameter allowed to vary is the tensor-to-scalar ratio, $r$.
All other parameters are fixed to the \planck best-fit values \citep{planck18_params}.
Expanding this parameter space to allow variation in other important parameters, such as the scalar amplitude $A_s$, is left to future work.
$EE$ and $BB$ spectra are computed for a scalar-only case with
lensing, and for a tensor-only case with $r=1$ and tensor spectral
index $n_t=0$.
The total CMB power is taken as the sum of the scalar modes with the tensor component scaled linearly with $r$.
This treatment therefore does not assume slow-roll Inflation.

The disadvantage of this approach to fitting parameters is that it is still limited by the approximations involved in the definition of the likelihood.
However, an important advantage is that it avoids the requirement for defining a bandpower likelihood for model comparison, and the approximations associated with that step.
Instead, direct evaluation of Equation~\ref{eq:DirectLikelihood} can be used in numerical searches for maximum likelihood parameter sets using MCMC techniques.
An additional advantage of this direct {\sl map-to-parameter} likelihood evaluation is that it becomes straightforward to marginalize the final parameter estimates over nuisance parameters such as noise calibrations or even foreground parameters, as we discuss below.

\section{Extensions}
\label{sec:extensions}

The XFaster likelihood-based approach can be extended in a number of ways in order to use the estimator for systematic checks, foreground reduction, and component separation.
These steps are an important part of any CMB analysis as they provide robustness and consistency checks, the ability to quantify systematic uncertainties, and the ability to determine the origin of any statistically significant signal.
The advantage of likelihood-based estimates is that they always yield an estimate of the Fisher matrix for the parameters.
This is useful for establishing the significance of any signal and for correctly propagating all uncertainties into the final estimates.

In this section, we show how XFaster has been adapted as an estimator for null test validation and foreground minimization.
In both cases, special consideration must be taken to properly treat sample variance, as the XFaster covariance construction assumes that all signal components are Gaussian random fields and thus susceptible to sample variance.
This is not appropriate for null tests, where differencing two maps removes the signal component.
Likewise, template-based foreground fitting relies on the assumption that the dust morphology is known, and thus sample variance is inappropriate to include.
Modifications to the pipeline to address these considerations are detailed below.

\subsection{Null Tests}
\label{sec:nulls}

Null tests are often used in CMB analyses to identify systematic noise biases and to evaluate the general quality of the data.
If undetected, any biases may be misinterpreted as signal.
The technique involves taking differences between disjoint subsets of the data, and comparing the residual spectra with that from simulations.
In this case, we take differences of maps produced using split-halves of the time-stream data.
There are a number of ways the splits can be defined to probe different potential systematic effects; the splits used for the \spider data set are described in \cite{bmode_paper}.
It is important to use a consistent estimation pipeline for each null split and for signal spectra.

With the XFaster estimator, null spectra can be evaluated using the same method one would use to calculate a signal power spectrum, but using the difference of two sets of maps, with some of the components handled differently.
The filter transfer function can only be estimated from ensembles where the signal is non-negligible; it is therefore computed using the ensemble average of the two simulated signal-only half maps. 
The model shape spectrum for null tests is flat and, once the spectrum has converged, the final Fisher matrix for a null spectrum is calculated without the sample variance component.
This is done by setting the final signal $q_b$ parameters to a very small value, thus nulling out the signal covariance.

The noise component term for null tests (appearing both in the covariance and data debias terms in Equation~\ref{eq:q_bestimator}) includes both signal and noise residuals, along with their correlations, as these terms all contribute to the expected variance and biases in the data spectra.
Unlike for total signal spectra, these terms are included for off-diagonal elements (cross-spectra of different half-maps) of the covariance as well; this is because all auto- and cross-spectra must account for expected residual signal and noise due to different filtering between the half-maps.

As an alternative to debiasing the data for expected signal and noise residuals in spectrum-space, the pipeline also has the option to subtract residuals with known morphologies in map-space.
When such maps are used, the covariance matrix noise term does not include signal contributions, since there is no sample variance in the debias term.
For example, to estimate signal residuals for the \spider null tests, we use \planck frequency maps processed and differenced in the same way as the data null maps, instead of using CMB signal simulations.
The frequency mismatch between the \spider and \planck bands could be accounted for, but was found to negligibly affect results.
We find that this properly accounts for foreground residuals due to slight differences in the time-domain filtering between the two halves, which dominate our null signal residuals at large scales for some data splits.
This method also allows us to accurately model the morphology of the residuals, and eliminates the need to account for sample variance from the subtraction in the covariance matrix.
The \planck-subtraction method has been tested with half-missions and half-rings, with negligible differences between the two.
This confirms that the residuals subtracted in this manner are dominated by signal rather than by noise in the \planck data.

The mode-counting factor $g_\ell$ is expected to be different for nulls in comparison to total signal spectra due to the different relative contribution of sample variance to the error (Figure \ref{fig:gcorr}).
Sample variance only affects the null test bandpower errors through its contribution to the uncertainty of the expected signal residual that debiases the data spectrum.
The mode counting factor cannot be empirically calibrated for null tests using signal-only simulations, because the remaining signal after debiasing with the expected signal spectrum for a signal-only simulation is nearly zero, which makes the covariance matrix singular and thus non-invertible.
Instead we add noise to the simulations, and calibrate $g_\ell$ iteratively in the same fashion as for total signal spectra.

We have found that the resulting $g_\ell$ is somewhat sensitive to the noise level used in its calibration.
Thus, we use the noise residual terms $n_b^i$ (Equation~\ref{eq:noise_component}) calculated for the data to rescale the $a_{{\ell}m}$ coefficients of the simulated noise maps as $\sqrt{(1+n_b^i)}\, a_{{\ell}m}^i$.
This modification affects both the $S{\times}N$ and $N{\times}N$ terms of the covariance matrices.

\subsection{Polarized Foreground Template Fitting}
\label{sec:template_fg}
On large angular scales, the polarized CMB signal is obscured by Galactic foregrounds.
The contribution from foregrounds biases any estimate of power on the sky with respect to the underlying cosmological signal.
This bias must be estimated and removed in order to recover the cosmological signal in a way that minimizes the impact on the final variance.
In \spider's observing region in both its \SIlist{95;150}{\giga\hertz} bands, the dominant foreground component is polarized dust \citep{bmode_paper}.
We therefore focus on dust in this section, though the method could also be adapted for other foreground components, such as polarized synchrotron emission.

There are a number of approaches that can be used to remove the foreground bias and include the effect of the removal in the error propagation.
If independent observations of the foreground signal exist, a template subtraction method can be used.
In the absence of accurate templates, both the estimate of foreground power and its subtraction from the data must be carried out {\sl internally}.
Here we described a template-based method implemented as part of the XFaster \spider analysis.

We model the dust with a map-space template to include the dust signal in our estimate of CMB bandpowers and parameter likelihoods.
We take advantage of the high signal-to-noise measurements of dust by the \planck instrument at \SIlist{217;353}{\giga\hertz} where dust is much brighter than the CMB \citep{Planck_CompSep}.
Because these maps also contain CMB, we subtract \planck's \SI{100}{\giga\hertz} measurement of the sky.
Since the \planck maps are calibrated against each other using the common CMB component at these frequencies, the residual in the difference map contains only foregrounds and noise.
We make the assumption that the morphology of the dust foreground does not depend on frequency, which is consistent with arcminute-resolution observations of dust polarized emission toward diffuse regions of the Milky Way between 353 GHz and 1.2 THz \citep{Ashton_2017}.
Under this assumption, the dust in the template map can be linearly scaled to match the dust in the \spider maps.
We label these linear coefficients for the two frequencies as $\alpha_{95}$ and $\alpha_{150}$.

Once the \planck templates are created, they are ``reobserved'' through the \spider pipeline such that they are subjected to the same filtering, beam, and cut-sky effects as the actual observations.
We then subtract $\alpha$-scaled templates from the data spectra, using \planck half-missions in each half of every cross-correlation so that no \planck auto-correlation noise enters the spectra.
The cleaned data spectra, $\widehat{\pmb{\Delta}}$, are given by
\begin{align}
\widehat{\pmb{\Delta}}^{ij} &= \left<(\pmb{m}_i - \alpha_i \pmb{t}_i) \times (\pmb{m}_j - \alpha_j \pmb{t}_j)\right> - \alpha_i\alpha_j\left<\pmb{n}_i^t \times \pmb{n}_j^t\right>\,,\nonumber\\
&= \widehat{\pmb{C}}^{i  j} - \alpha_i \left<\pmb{t}_i \times \pmb{m}_j\right> - \alpha_j \left<\pmb{m}_i \times \pmb{t}_j\right> + \alpha_i \alpha_j \left<\pmb{t}_i \times \pmb{t}_j\right> \nonumber\\
& \hphantom{=} \, - \alpha_i\alpha_j\left<\pmb{n}_i^t \times \pmb{n}_j^t\right>\,,
  \label{eq:template_clean}
\end{align}
where the $\left<\ldots\times\ldots\right>$ indicate cross pseudo-spectra of two maps, $i$ and $j$ are map indices, $\pmb{m}$ represents a \spider map, $\pmb{t}_i$ and $\pmb{t}_j$ are different half-mission templates reobserved to match the \spider map, $\pmb{n}_i^t$ and $\pmb{n}_j^t$ are different reobserved half-mission template noise simulations, and $\alpha$ values are the linear scaling factors.
Each of the terms in Equation \ref{eq:template_clean} is computed once and subsequently scaled with $\alpha$ values that are varied with each iteration in the likelihood.

We have chosen to subtract the scaled template from the data rather than to add it to the model covariance since the latter method would add unnecessary sample variance to the covariance.
To account for the error introduced from the template subtraction, such as from \planck noise or chance correlations between the CMB and the template, we run the algorithm on an ensemble of simulations.

The simulations include CMB signal realizations, \spider noise, a foreground template (either directly using a \planck template or a scaled Gaussian realization of a power law), and \planck noise maps from the FFP10 simulation ensemble\footnote{\planck end-to-end ``full focal plane'' simulations \citep{planck18_hfi}}.
The distribution of parameters determined for this ensemble is taken to be the true covariance.
The average covariance the XFaster algorithm estimates is the total covariance without error from template fitting.
The difference of the two is taken to be the additional contribution to the covariance due to the foreground subtraction, and is added to each Monte Carlo sample in the data parameter likelihoods.

The amplitude of this additional covariance is found to be independent of simulated $r$ and foreground morphology.
However, it scales with $\alpha$, as expected since $\alpha$ scales the \planck noise contribution.
Because of this, we compute the term with the precise data-preferred $\alpha$ values applied to the simulations.

The \planck FFP10 noise simulations are correlated with each other by a non-negligible amount, due to the method by which they were produced.  To account for these correlations, we include the scaled average of \planck half-mission 1$\times$half-mission 2 noise in the terms subtracted from the data.

\subsection{Harmonic-Domain Foreground Fitting}
\label{sec:harmonic_fg}
An alternative method of foreground fitting involves estimating the contribution of frequency-dependent foregrounds without requiring a map-space template.
This method necessarily relies on an assumed model for the frequency dependence of the contributing foreground.
The model can be included in either the map estimation step as a contribution to the model for the observed data or in the power spectrum estimation step as a contribution to the sample variance in the likelihood. Here we describe an extension to XFaster using the latter.

We model the Spectral Energy Density (SED) of the dust component as
\begin{equation}
S_d(\nu) = A_d \frac{B_{\nu}(T_d)}{B_{\nu_0}(T_d)}\left(\frac{\nu}{\nu_0}\right)^{\beta_d-2}\,,
\end{equation}
where $T_d$ is the blackbody temperature of the dust, $A_d$ is the model amplitude at reference frequency $\nu_0$, $B_\nu(T)$ is the blackbody spectrum at temperature $T$, and $\beta_d$ is a spectral index.
The SED describes the brightness temperature of the dust.
To relate this to the representation of CMB maps using thermodynamic temperature $\Theta$, in which blackbody sources are frequency independent, we use the idealized conversion factor
\begin{equation}
\Theta_d(\nu) = \frac{(e^x-1)^2}{x^2e^x} S_d(\nu) \equiv g(\nu) S_d(\nu)\,,
\end{equation}
where $x = h\nu/k T_{CMB}$.
In practice the $g(\nu)$ factors must be corrected for the specific frequency dependence of experimental window functions.
These factors are color-corrected for the \spider data by integrating over the \SIlist{95;150}{\giga\hertz} window functions \citep{shaw_spie}.

The dust contribution to the model, cut-sky covariance (Equation~\ref{eq:signal}) is given by
\begin{equation}
{\widetilde S}_{\ell}^{d,ij} = \frac{\Theta_d(\nu_i) \Theta_d(\nu_j)}{\Theta^2_d(\nu_0)} \sum_{(dust)\,b} q^d_b \widetilde C_{b\ell}^{d,ij}\,.
\end{equation}
We have introduced a set of dust bandpower parameters $q_b^d$ with $b$ a set of bands defined specifically for the dust component.
We use bandpower kernels $\widetilde C_{b\ell}^{d,ij}$---including all kernel, filter, and beam terms as in Equation~\ref{eq:cbl_tt}---in which we parameterize the full-sky dust shape spectrum as $\ell (\ell + 1) C^{d\,(S)}_\ell / 2\pi = A (\ell/80)^{\alpha+2}$.
The amplitude $A$ and angular spectral index $\alpha$ are set to the best-fit values reported in \cite{Planck_CompSep}.
The bandpower parameters $q_b^d$ are varied simultaneously with all other bandpower parameters to find a global, maximum likelihood fit for the combination of \emph{e.g.}, CMB, dust, and noise residual bandpowers for each polarization combination.
Since the dust bandpowers appear at linear order in the covariance,
the estimator in Equation~\ref{eq:q_bestimator} is unchanged as long
as other parameters such as $T_d$, $\beta_d$ are fixed. If the assumed
values for the additional parameters are not correct, the effect will
be incorporated into the bandpower parameters $q_b^d$, and these can be
regarded as the ``effective'' rescalings of the model.\footnote{An
  alternative is to include a non-linear maximization of the
  likelihood over, \emph{e.g.},
  $\beta_d$ at each step of the quadratic iteration.}
We explore the use of harmonic space foreground fitting in future work.

\section{Pipeline Validation}
\label{sec:validation}
To validate the pipeline, we conduct a series of tests with simulated inputs that each require two criteria to be met.
\begin{enumerate}
\item The ensemble averaged parameter estimates produced by XFaster must match the input values used to generate the simulations.
\item The error in the estimate must match the scatter of individual XFaster estimates over the ensemble.
\end{enumerate}
These conditions, achieved to within set tolerance levels, ensure that the XFaster estimates are unbiased in both mean and variance.
In practice, any recalibration of the Fisher matrix required to satisfy the second condition minimizes the effect of the approximation used to define the XFaster likelihood.
This is achieved at the cost of some additional simulations used for the effective calibration of the likelihood for specific data sets.
The pipeline must also behave well for reasonable changes to the simulation inputs, such as different cosmological parameters, foreground morphologies, or noise amplitudes within expected ranges for the data.

In this section, we first describe the simulation inputs, followed by the results of validation tests for each of the pipeline outputs.

\subsection{Simulated Maps}
Simulation ensembles include a CMB signal component, an instrument noise component, and (optionally) a foreground signal component.
Each of these are described in detail in the following sections.

\subsubsection{CMB Maps}
CMB maps are produced using the \texttt{synfast HEALPix} routine \citep{HealPix}, which generates Gaussian realizations of input angular power spectra.
The power spectra are computed by the software package \texttt{CAMB} \citep{camb} from a fiducial set of $\Lambda$CDM best-fit parameters to the \planck data \citep{planck15_params}.
To generate appreciable signal for computing the transfer function, we construct the simulation ensemble by replacing the fiducial $\Lambda$CDM input $BB$ spectrum with a spectrum that is flat in $\ell(\ell+1) C_\ell$.

Each signal realization is generated at \texttt{HEALPix} resolution
$N_\mathrm{side}=2048$, then smoothed with a \spider beam per focal plane.  Each
realization of Stokes $T$, $Q$ and $U$ maps is then ``reobserved'' through the
\spider mapmaking pipeline, applying all flagging and time-domain filtering
identically to the measured data.  The $T$ component of each input realization
is then passed through the filtering and mapmaking pipeline a second time.  The
resulting $Q$ and $U$ maps then contain only the temperature-to-polarization
leakage induced by the pipeline, and are subtracted from the first set of
reobserved $Q$ and $U$ maps to remove any bias due to this leakage.

This leakage-subtraction procedure was deemed prohibitively costly for null tests, since signal simulations must be made for each of ten null splits.
Instead, the null signal simulations use CMB maps with the $T$ map replaced with a \planck map at the frequency closest to its \spider analog, and the $E$-mode power constrained by the $TE$ cross-correlation spectrum.
This means the $T$-to-$P$ leakage map is the same for every seed in the signal ensemble.
Because null tests use the difference of maps with approximately the same leakage correction, this further approximation to the leakage correction negligibly affects results.

We use 1000~unconstrained CMB realizations for computing bandpowers and
likelihoods, and 500~constrained CMB realizations for computing bandpowers for
each of ten null splits.  To preserve any data correlations between null tests
that are present in the measured data, we use the same set of random number
generator seeds for the signal components going into each null split.

\subsubsection{Noise}
Noise simulations are generated in the time domain by sampling from a single power spectral density (PSD) per channel.
The input PSD is constructed by averaging over the entire flight in ten-minute chunks; in practice, this overestimates the \spider noise due to the asymmetric impact of high outliers in the distribution.
As with the signal simulations, the noise realization for each channel uses the same set of random number generator seeds for each null split, in order to preserve noise correlations that are similarly present in the data.

\subsubsection{Foregrounds}
We require simulated foreground templates to test the template cleaning method of foreground separation.
Each simulated template is constructed from a fiducial foreground template with added instrument noise from the \planck FFP10 simulation ensemble, where both the templates and noise realizations are reobserved with the \spider mapmaking pipeline.
The foreground template ensemble is limited by size of the FFP10 data set to 300~such realizations.
The fiducial foreground template is constructed by differencing two \planck frequency maps as usual, \emph{e.g.}, \planck \SI{353}{\giga\hertz} minus \SI{100}{\giga\hertz}.
The fiducial template also includes \planck instrument noise by construction, but it is our best estimate of the morphology of the true foregrounds.
Alternatively, we can construct the fiducial template using a Gaussian realization of a power law matching the \planck best-fit dust power law model.
We find these two foreground simulation methods behave equally well in the validation results.

\subsection{Bandpower Validation}

\begin{figure}
  \centering
  \includegraphics[width=\columnwidth]{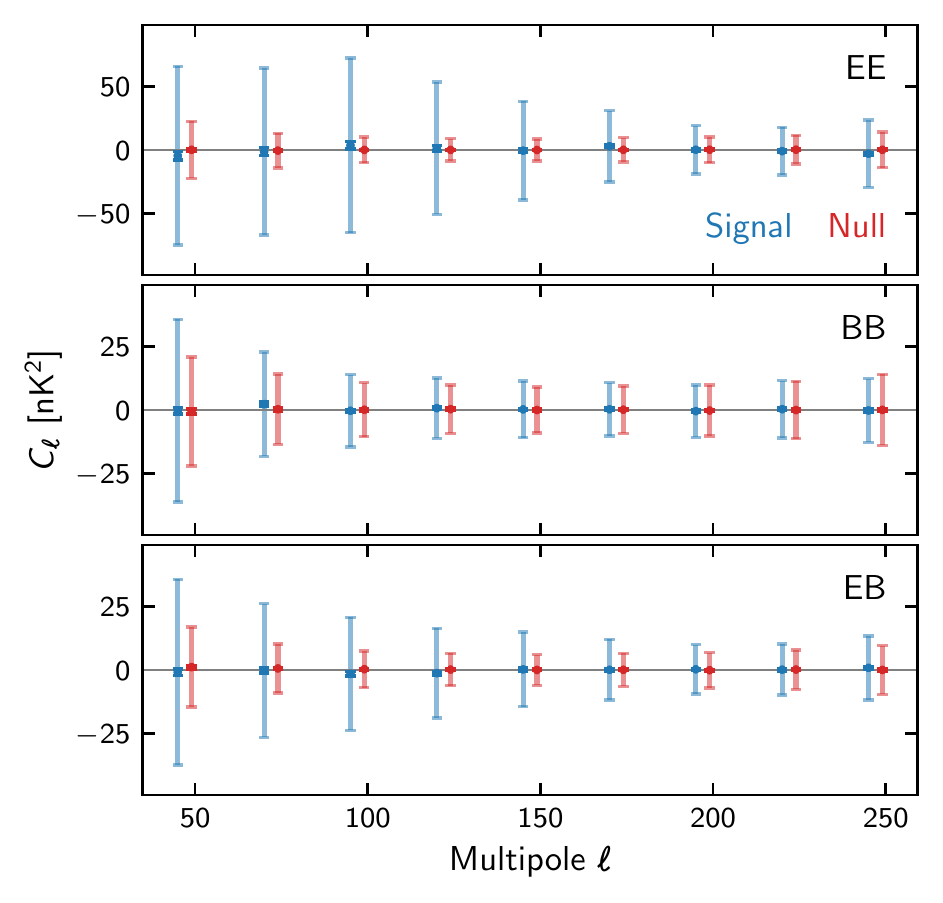}
  \caption{Residual spectra averaged over an ensemble of 500~signal (\emph{blue}) and null test (\emph{red}) simulations differenced with the corresponding input model spectrum.
    Dark error bars show the error on the mean.
    Light error bars indicate the total spectrum error.
    Sample variance is included for signal spectra and excluded for null spectra.
    The null test shown is the checkerboard split (see \cite{bmode_paper} for null split definitions), though all other splits show similar results.
    }
  \label{fig:bandpower_bias}
\end{figure}

To validate the XFaster bandpower estimation pipeline, we compute bandpowers for an ensemble of 500~CMB+noise simulations for both signal and null spectra.
We verify that the average of the computed signal bandpowers matches the spectrum input to generate the CMB maps and is therefore unbiased, and that the average of the null bandpowers is consistent with zero.
These bandpower distributions are shown in Figure~\ref{fig:bandpower_bias}.

\begin{figure}
  \centering
  \includegraphics[width=\columnwidth]{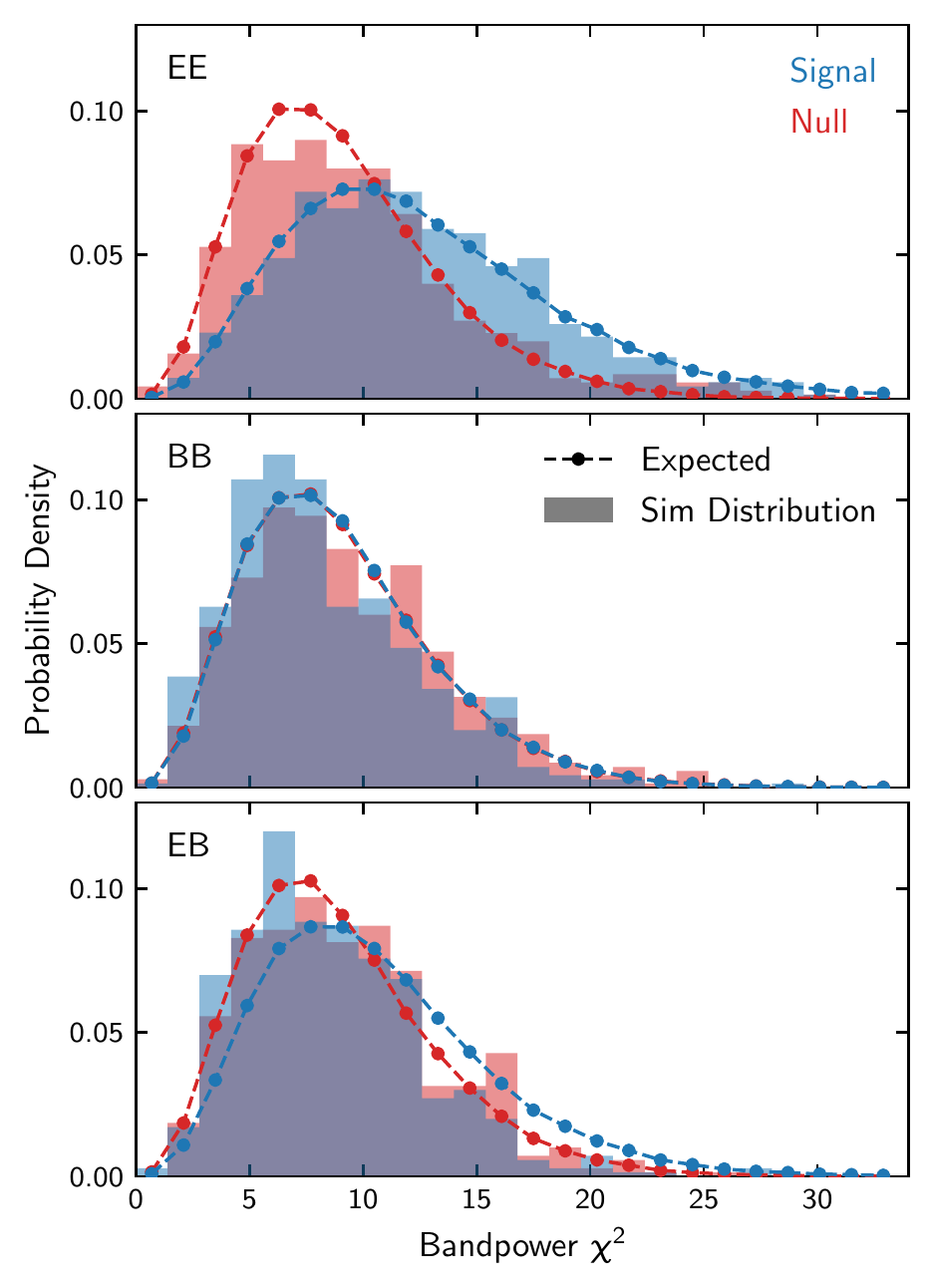}
  \caption{Distributions of bandpower $\chi^2$ estimates for an ensemble of 500~simulations (\emph{histograms}) compared to expectations from the corresponding covariance (\emph{dotted lines}).  Results are shown separately for CMB signal simulations (\emph{blue}) and the checkerboard null split (\emph{red}), which is representative of the other nine null splits.
    $\chi^2$ values are calculated for nine multipole bins extending from $33<\ell<257$, using a fiducial $\Lambda$CDM model for the signal ensemble, and a null model for the null split ensemble.
    The expectation distribution is determined by generating 100,000 realizations of bandpowers from the Fisher covariance matrix, and histogramming the resulting $\chi^2$ values.
    Agreement between the histogram and expectation lines indicates that the covariance matrix is accurate.
    The covariance is slightly overestimated in $BB$ and $EB$ signal, producing lower $\chi^2$ values than expected from random realizations of the covariance matrix.}
  \label{fig:chi2_bandpowers}
\end{figure}

To verify that the covariance computed by XFaster is accurate, we check that the
scatter of the ensemble matches the covariance computed by XFaster for both
signal and null simulations.  We construct two distributions of $\chi^2$ values:
\begin{enumerate}
\item The $\chi^2$ of the spectrum for each of 500~realizations relative to the
  corresponding input model (fiducial $\Lambda$CDM for the signal ensemble, or a
  null model for the null ensemble).
\item The $\chi^2$ of each of 100,000 spectrum realizations, sampled from the
  average of the bandpower covariance matrices computed for each of the
  realizations in the signal and null map ensemble.
\end{enumerate}
The second distribution forms the expectation for the first distribution; as shown in Figure~\ref{fig:chi2_bandpowers}, the two distributions are in good agreement.
The covariance is slightly overestimated in $EE$ and $BB$ for signal spectra, indicating that error bars for these data spectra will be somewhat larger than is optimal.
These results are insensitive to input signal shape and noise amplitude.

\subsection{Likelihood Validation}

\begin{figure}
  \centering
  \includegraphics[width=\columnwidth]{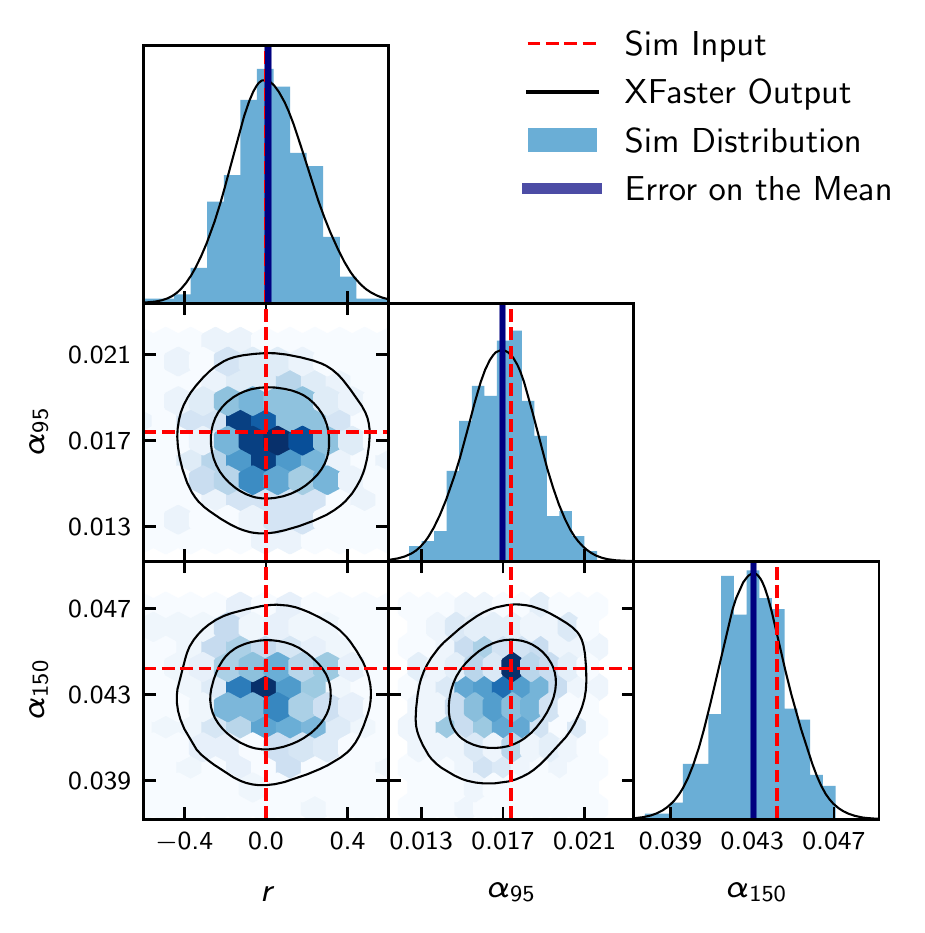}
  \caption{XFaster likelihood results for an ensemble of 300 CMB+noise+foreground simulations.
    Red dotted lines indicate the input $r$ and foreground template scalings, $\alpha_{95}$ and $\alpha_{150}$, used to make simulated maps.
    The XFaster likelihood contours computed for the ensemble mean of the simulations are shown in black, and the maximum likelihood
    parameter values computed for each realization in the ensemble are histogrammed in blue, with dark blue shading to indicate the mean and error on the mean of the distribution.
    The agreement between the histograms and the black contours shows that the XFaster pipeline is unbiased in the mean and in its estimate of error for $r$.
    The biases in $\alpha$, most evident at \SI{150}{\giga\hertz}, are small compared to the error and do not affect the cosmological result.
    }
  \label{fig:like_validation}
\end{figure}

The likelihood pipeline was tested on ensembles of CMB+noise+dust ``fake
data'' maps, and dust+\planck noise ``fake template'' maps.
The parameters $r$, $\alpha_{95}$, and $\alpha_{150}$ were varied
using a Monte Carlo Markov Chain sampler, and
all other parameters were fixed to the \planck best-fit values from Table 1 in \cite{planck18_params}.
The pipeline was deemed to be validated if the output parameter likelihoods were unbiased with respect to the input parameters, and if the likelihood widths matched the scatter of the best-fit parameters from the ensemble.
The pipeline was tested using various values of $r$, ranging from 0 to 0.7, various noise amplitudes, and various template morphologies.
In addition, we fit for and marginalized over beam uncertainty and noise residual amplitudes; these nuisance parameters were found to have a negligible effect on the data posteriors for a \spider-like data set.
The results for the nominal simulation data set are shown in Figure~\ref{fig:like_validation}.
The output parameter likelihoods are not biased with respect to the input parameters, and the likelihood widths match the scatter of the best-fit parameters from the ensemble, indicating that the XFaster likelihood is accurate and unbiased.

The recovered $\alpha_{95}$ and $\alpha_{150}$ parameters indicate a slight bias in the estimator, as evidenced by comparing the 1$\sigma$ error on the the distribution mean to the simulation input values.
However, these biases are small with respect to the error on the parameters, and evidently do not impact the cosmological result, as the recovered maximum likelihood $r$ value is within 1$\sigma$ error on the mean (0.011) of the input $r$ to the simulation.

\section{Application to \spider Data}
After extensive validations of the XFaster pipeline using simulations, the analysis was applied to the data from the 2015 flight of the \spider instrument.  The comprehensive results of this analysis, using XFaster and additional pipelines, are presented in \cite{bmode_paper}.
A subset of the XFaster results are reproduced here as a demonstration of the pipeline functionality.
\label{sec:spiderdata}

\subsection{Null Tests}
\label{sec:result_null}
\begin{figure}
  \centering
  \includegraphics[width=\columnwidth]{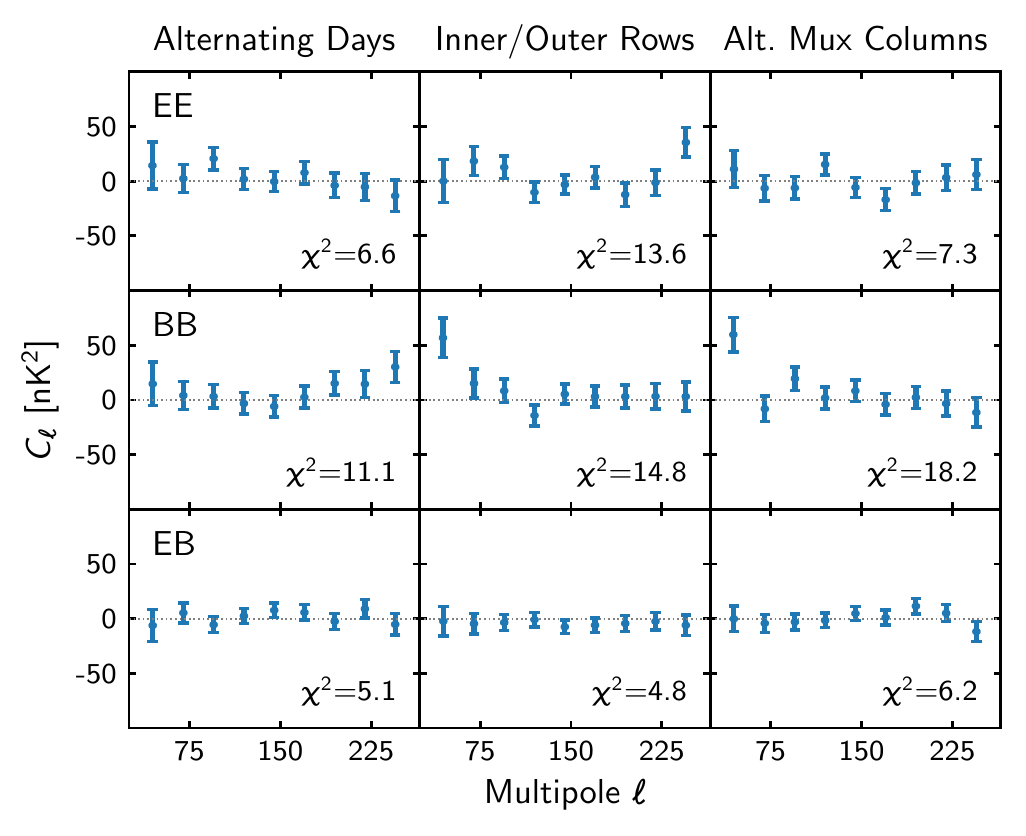}
  \caption{Three \spider null tests, showing $EE$, $BB$, and $EB$ spectra for the combined \SIlist{95;150}{\giga\hertz} data, with $\chi^2$ values computed for the bins shown.}
  \label{fig:spider_nullspectra}
\end{figure}

Ten different null tests were carried out on the \spider data set, including seven detector-based splits and three time-based splits.
Nine of the splits were designed to probe potential systematic errors in the data; one, the checkerboard pattern detector split, was chosen to be largely insensitive to systematics, instead probing the pipeline's handling of the noise model.
The full list of splits is detailed in \cite{bmode_paper}.
The estimated null spectra for some representative subsets of data splits are shown in Figure \ref{fig:spider_nullspectra}.

By construction, different null splits contain overlapping detector samples and therefore are correlated at some level.
For example, the \spider ``inner rows'' null half overlaps by 75\% with the ``inner radius'' null half.
This creates a challenge in assessing statistics of the full null ensemble.
These correlations are preserved among simulated maps generated using the same random number generator seed, so we can use simulations to inform our expectations for distribution shape statistics.

\begin{figure}
  \centering
  \includegraphics[width=\columnwidth]{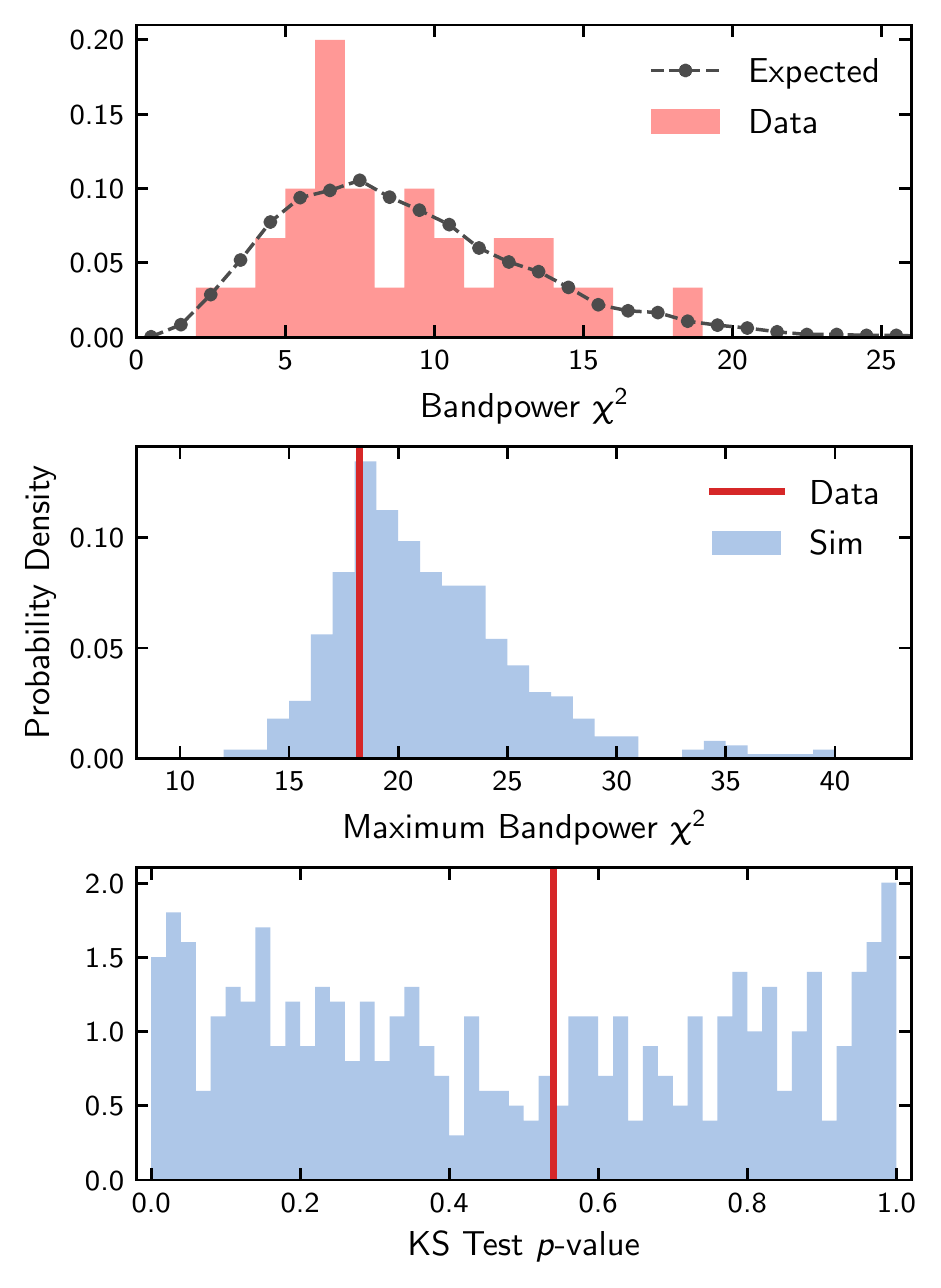}
  \caption{Illustration of null statistics computed from the combined \spider data set.  (\textit{top}) Histogram of all null test bandpower ${\chi}^2$ values and expected distribution of drawn ${\chi}^2$s from the covariance matrix.
    The red histogram consists of 30 ${\chi}^2$ values, (3 polarization spectra for 10 tests).
    The expectation histogram (\textit{dashed line}) consists of 5000 random bandpower draws from each covariance matrix for each test, used to compute a total of 150,000 ${\chi}^2$s.
    A KS test is performed between the data and expectation distributions, resulting in a $p$-value of 0.54.
    (\textit{middle})
    Histogram of the maximum bandpower $\chi^2$ value from 500 end-to-end simulations. The maximum $\chi^2$ of the data is 18.2, corresponding to a PTE of 0.78 given the ensemble of max-$\chi^2$ values. 
    (\textit{bottom}) Histogram of the KS test $p$-values from the same set of 500 end-to-end simulations.  The $p$-value of the data (\textit{red line}) corresponds to a PTE of 0.55 given the ensemble of KS tests. Together, these tests indicate that the combined \spider data set passes its suite of null tests.
    }
  \label{fig:spider_nullstats}
\end{figure}

When such correlations are negligible, the expected distribution of the null ${\chi}^2$ values can be determined by drawing simulated null bandpowers from the covariance matrix.
We can then perform a Kolmogorov-Smirnov (KS) test that computes the likelihood that the data is drawn from the cumulative distribution extrapolated from the simulated ${\chi}^2$ values.
The distributions used in this test for the \spider combined data set are shown in the top
panel of Figure~\ref{fig:spider_nullstats}; the resulting KS test $p$-value is 0.54, indicating good agreement.

To account for the presence of correlations, we then perform the same test on each of 500 end-to-end simulations.
We compare the KS test $p$-value from data to the distribution of the simulation KS test $p$-values to test how unlikely our data is relative to simulations.
These results are shown in the bottom panel of Figure~\ref{fig:spider_nullstats}, indicating that the \spider data exceed the criterion of being at least 1\% likely given the simulation ensemble.
We further test the outliers of the data distribution by comparing the maximum data ${\chi}^2$ to the maximum ${\chi}^2$ in each end-to-end simulation.
We again find that the outlier data ${\chi}^2$s are more than 1\% likely, as shown in the middle panel of Figure~\ref{fig:spider_nullstats}.

The flexibility, and reduced computational overhead, introduced by the XFaster approximate likelihood method is a key advantage of the pipeline. It allows a full explorations of null testing splits, internal calibration of systematics and determination of goodness-of-fit statistics for any estimated quantity.

\subsection{Bandpower Estimates}

The XFaster procedure applied to the \spider data set requires a set of $\sim 100$ parameters, including both signal and noise components, and as many as $\sim 300$ parameters when the harmonic-domain foreground model is also included.
We construct the XFaster estimator to compute 16 bandpowers, equally spaced between $\ell=8$ and $\ell=407$ for each of six polarization spectra, with additional noise calibration residual bandpowers defined over the same bins.
We have found the choice of noise bin width does not significantly affect the result, as the \spider noise model is not very degenerate with the expected signal.
The nine bins between $\ell=33$ and $\ell=257$ are used for subsequent analyses; the other seven bins are included in the estimator to accurately account for their contribution to the so-called science bins through bin-to-bin leakage.
All spectra contain information from modes at $\ell \lesssim 8$ that are so heavily correlated due to the reduced sky coverage that the XFaster approximation breaks down, even after calibration of the effective mode count.
To reduce the sensitivity to these correlations we exclude all modes at $\ell < 8$ in all spectral decompositions. 
The second bin ($8\leq \ell<33$) is included in the iterative bandpower estimation procedure, but discarded in later analysis steps due to the level of correlation with the unconstrained lowest multipoles.

Convergence of the XFaster estimator is reached at iteration $i$ when
\begin{equation}
  \max_b \left| \frac{\theta_b^{(i)} - \theta_b^{(i-1)}}{\theta_b^{(i-1)}} \right| < 0.005\,,
\end{equation}
\emph{i.e.}, when all model parameters change by less than 0.5\% relative to their values at the previous iteration.
This requires ${\mathcal O}(10)$ iterations of the quadratic estimator in Equation~\ref{eq:q_bestimator}.

\begin{figure}
  \centering
  \includegraphics[width=\columnwidth]{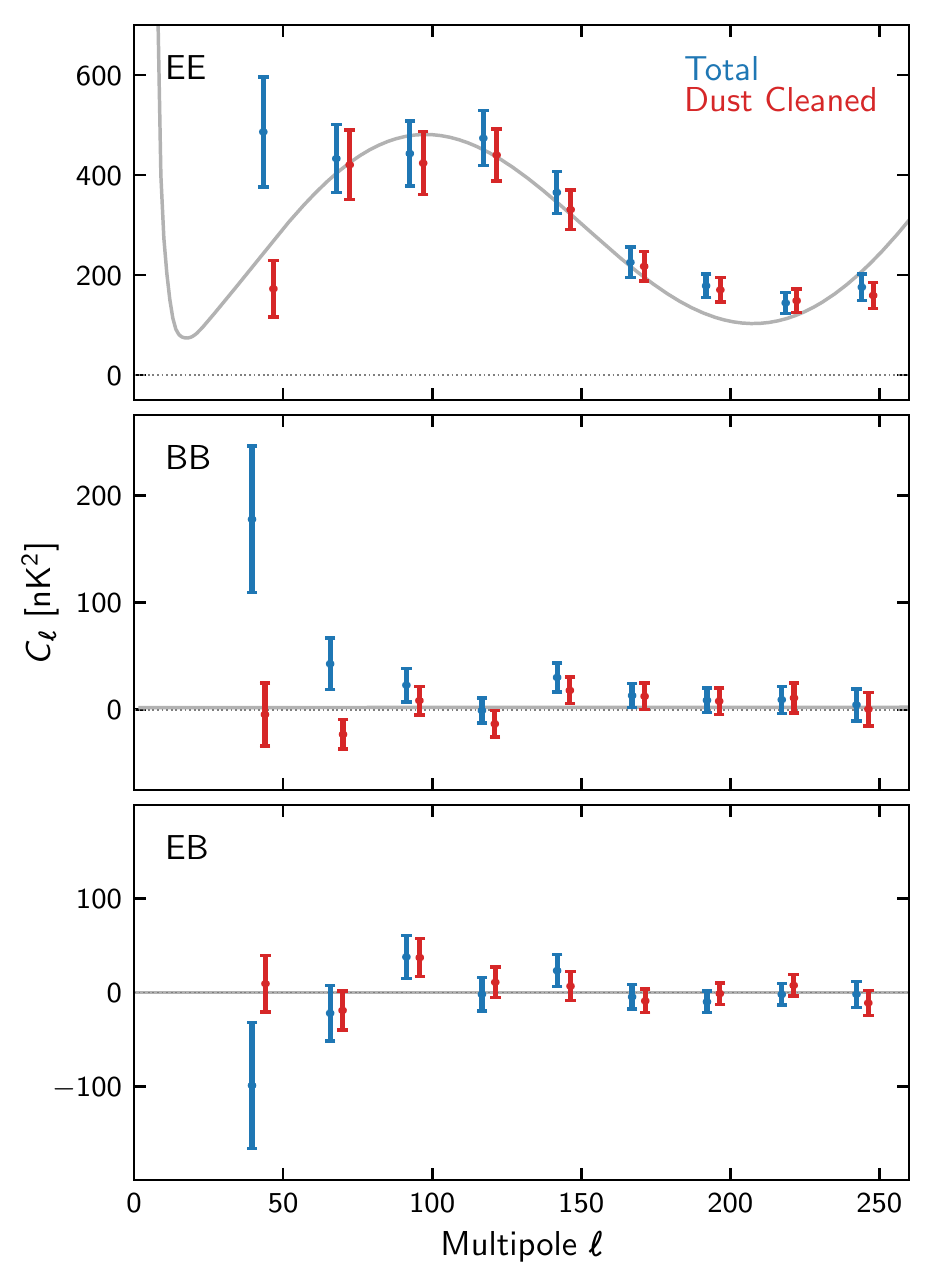}
  \caption{\spider estimated bandpowers estimates for total power (blue) and CMB only (red), computed from a combination of \SI{95}{\giga\hertz} and \SI{150}{\giga\hertz} data.
    Sample variance of the total estimated power is included, and foreground-cleaned error bars include error from template-fitting.
    A fiducial $\Lambda$CDM CMB spectrum with $r=0$ is shown in gray.}
  \label{fig:spider_bandpowers}
\end{figure}

Figure \ref{fig:spider_bandpowers} shows the estimated bandpowers computed from the raw \spider data set (\emph{i.e.}, making no attempt to remove any foreground contamination), as well as the bandpowers obtained using the template-based dust subtraction method of Section~\ref{sec:template_fg}.
The bandpower error bars are obtained from the diagonal of the bandpower covariance (Equation~\ref{eq:cov_from_fisher}), computed from the inverse of the Fisher matrix after discarding all nuisance parameters (noise residuals in this case).
Spectra are computed using a data set containing both \SIlist{95;150}{\giga\hertz} observations with a common sky mask.
The results show significant dust power in the raw spectra at large angular scales.
Dust-cleaned spectra are in good agreement with expectations from $\Lambda$CDM, and the contribution to the angular power spectrum from inflationary gravitational waves, parameterized by $r$, is not detected.

\subsection{Likelihood Estimates}

\begin{figure}
  \centering
  \includegraphics[width=\columnwidth]{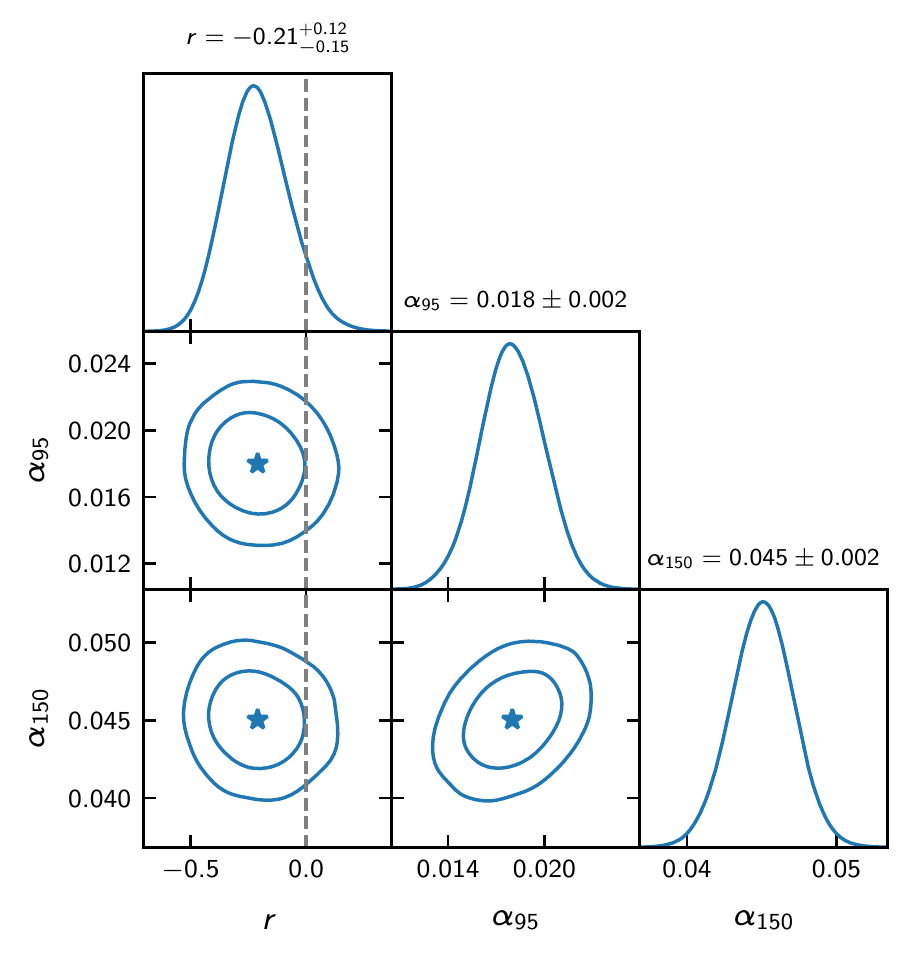}
  \caption{The combined XFaster likelihood for $r$ and $\alpha$, imposing no priors on these parameters and using a \planck $353-\SI{100}{\giga\hertz}$ template to estimate the foreground morphology.
    1$\sigma$ constraints are shown in the panel titles.}
  \label{fig:spider_likelihoods}
\end{figure}

Finally, we demonstrate the application of XFaster as a direct map-to-parameter
likelihood estimator for the \spider data set.  The likelihood model includes
foreground templates constructed from the \planck $353-\SI{100}{\giga\hertz}$
difference map, parameterized with \SIlist{95;150}{\giga\hertz} scalings,
$\alpha_{95}$ and $\alpha_{150}$; a CMB signal component parameterized by the
tensor-to-scalar ratio $r$ and all other $\Lambda$CDM parameters held fixed; and
rescaled noise components as determined by the iterative fitting procedure.  The
noise model parameters are marginalized over for the final result, along with
uncertainties in the beam model; these nuisance parameters do not contribute
significantly to the uncertainties on the cosmological and foreground
parameters.  Profile likelihoods in the directions along $r$, $\alpha_{95}$ and
$\alpha_{150}$ are shown in Figure~\ref{fig:spider_likelihoods}.

Further discussion of this result, including construction of an appropriate upper
limit on the the tensor-to-scalar ratio $r$, can be found in \cite{bmode_paper}.

\begin{table*}
  \centering
  \caption{Pipeline steps and computing requirements, along with their order of magnitude scaling for relevant pipeline parameters.
      Fiducial numbers use eight $N_\mathrm{side}=512$ maps, $\ell_\mathrm{max}=407$, and a single core.
      Where changing these quantities affects the time or memory requirements, they are listed with their approximate scaling.
      $N_\mathrm{map}$ is the number of data maps, $N_\mathrm{x-spec}=N_\mathrm{map}(N_\mathrm{map}+1)/2$ is the number of cross spectra, $N_\mathrm{pix}$ is the number of pixels in each map, $N_\mathrm{sim}$ is the number of simulations (assumed to be the same for signal and noise), $\ell_\mathrm{max}$ is the maximum multipole used, and $N_\mathrm{param}$ is the number of parameters solved for in the likelihood.}
  \label{table:compute_req}
  \begin{tabular}{lrrrrr}
    \toprule
    \toprule
    \textbf{Pipeline Step} & \textbf{CPU Time} & \textbf{Time Scaling} & \textbf{Memory (GB)} & \textbf{Memory Scaling} & \textbf{OMP Speed Up}\\
    \midrule
    Mask cross spectra & 1 min & $N_\mathrm{x-spec}$, $N_\mathrm{pix}, \ell_\mathrm{max}^2$ & 1.3 & $N_\mathrm{map}$, $N_\mathrm{pix}$, $\ell_\mathrm{max}$ & Minimal\\
    Mode coupling kernel & 6 min & $N_\mathrm{x-spec}$, $\ell_\mathrm{max}^2$ & 0.75 & $N_\mathrm{x-spec}$, $\ell_\mathrm{max}^2$ & None\\
    Pseudo-spectra of simulated maps & 16 hr & $N_\mathrm{sim}$, $N_\mathrm{x-spec}$, $N_\mathrm{pix}$, $\ell_\mathrm{max}^2$ & 2.2 & $N_\mathrm{map}$, $N_\mathrm{pix}$, $\ell_\mathrm{max}$ & $\sqrt{CPU}$\\
    Filter transfer function & 10 s & $N_\mathrm{x-spec}$, $\ell_\mathrm{max}^2$ & 0.8 & $N_\mathrm{x-spec}$, $\ell_\mathrm{max}^2$& None\\
    Pseudo-spectra of data maps & 100 s & $N_\mathrm{x-spec}$, $N_\mathrm{pix}$, $\ell_\mathrm{max}^2$ & 1.3 & $N_\mathrm{map}$, $N_\mathrm{pix}$, $\ell_\mathrm{max}$ & $\sqrt{CPU}$\\
    Bandpowers & 3 min & $N_\mathrm{x-spec}$, $\ell_\mathrm{max}^2$ & 1.9 & $N_\mathrm{x-spec}$, $\ell_\mathrm{max}^2$& Minimal\\
    Likelihoods & 5 hr & $N_\mathrm{param}$, $N_\mathrm{x-spec}$, $\ell_\mathrm{max}$ & 1.9 & $N_\mathrm{param}$, $N_\mathrm{x-spec}$, $\ell_\mathrm{max}$ & None\\
    \bottomrule
  \end{tabular}
\end{table*}

\section{Using the Code}
\label{sec:code}
The pipeline is written entirely in Python and is available on GitHub.\footnote{\href{https://github.com/annegambrel/xfaster}{https://github.com/annegambrel/xfaster}}
While it is possible to run the code on a single processor, it is greatly sped up by parallelizing matrix operations with OpenMP.
Further speed gains could be achieved by using MPI or high-throughput computing to distribute pseudo-spectrum computation across multiple processors, but this functionality is not currently implemented.
The code base also includes tools for submitting individual jobs on computing clusters using the Slurm workload manager.\footnote{\href{https://slurm.schedmd.com/}{https://slurm.schedmd.com/}}

The most time-consuming step in the analysis process is computing the pseudo-spectra of each of the simulated map crosses using the \texttt{anafast} method of the \texttt{healpy} package.
However, this step only needs to be performed once per mask choice, as its results are stored to disk and read in for bandpower and likelihood computations.
The XFaster method applied to the \spider data set included 1000 simulations, $\ell_\mathrm{max}=407$, $N_\mathrm{side}=512$, and eight independent maps (four at \SI{95}{\giga\hertz} and four at \SI{150}{\giga\hertz}), which required approximately 5~hours on 20~cores and 2~GB of memory to complete the bandpower computation from the individual maps on disk.
By contrast, the longest step of the full \spider analysis pipeline---generating the ensemble of 1000 signal and noise simulation maps---requires approximately 45 core-years.

The sequential steps of the XFaster algorithm are listed in Table \ref{table:compute_req} along with the time and memory required for each step.
Intermediate results are written to disk after each step is completed, so that subsequent steps may be run starting from that checkpoint.  For example, it is typical to construct the mask and simulation ensemble spectra once, then adjust bins and nuisance parameters for the particular application.
Due to the modular structure of the code, the total run-time required is the sum of the rows (modified by the number of CPUs provided for each step), and the memory required is simply the maximum among the rows, or 2.2~GB for the \spider fiducial case.

\section{Conclusion}
\label{sec:conclusion}
We have presented the XFaster power spectrum and parameter likelihood estimation package, and have demonstrated its validation with simulations and application to the \spider 2015 data set.
XFaster builds upon the MASTER formalism for estimating full sky CMB power spectra and covariances, accounting for filtering and noise biases using an ensemble of simulations.
Unlike the MASTER method, it estimates the covariance of bandpowers using an iterative calibration of the Fisher matrix, and therefore only requires one set of signal and noise simulations that do not need to be precise representations of the data.
The result is a pipeline that can produce fast, accurate power spectra and likelihoods for cosmological parameters.
It is additionally capable of computing null spectra and fitting for Galactic foregrounds, all within a self-consistent, self-contained framework.
This pipeline has been thoroughly validated with simulations of the \spider data set, and is publicly available for use on other CMB data sets.

\section*{Acknowledgments}
We acknowledge the contribution to the development of the XFaster
pipelines by all members of the BOOMERanG, \planck, and \spider collaborations.
\Spider is supported in the U.S. by the National Aeronautics and Space Administration under grants NNX07AL64G, NNX12AE95G, and NNX17AC55G issued through the Science Mission Directorate and by the National Science Foundation through PLR-1043515.
Logistical support for the Antarctic deployment and operations is provided by the NSF through the U.S. Antarctic Program.
Support in Canada is provided by the Natural Sciences and Engineering Research Council and the Canadian Space Agency.
Support in Norway is provided by the Research Council of Norway.
Support in Sweden is provided by the Swedish Research Council through the Oskar Klein Centre (Contract No.\ 638-2013-8993) as well as a grant from the Swedish Research Council (dnr.\ 2019-93959) and a grant from the Swedish Space Agency (dnr.\ 139/17).
The Dunlap Institute is funded through an endowment established by the David Dunlap family and the University of Toronto.
The multiplexing readout electronics were developed with support from the Canada Foundation for Innovation and the British Columbia Knowledge Development Fund.
AEG is supported by the Kavli Institute for Cosmological Physics at the University of Chicago through an endowment from the Kavli Foundation and its founder Fred Kavli.
CRC was supported by UKRI Consolidated Grants, ST/P000762/1,
ST/N000838/1, and ST/T000791/1.
KF holds the Jeff \& Gail Kodosky Endowed Chair at UT Austin and is grateful for that support.
WCJ acknowledges the generous support of the David and Lucile Packard Foundation, which has been crucial to the success of the project.

Some of the results in this paper have been derived using the \texttt{HEALPix} package \citep{HealPix}.
The computations described in this paper were performed on four computing clusters: Hippo at the University of KwaZulu-Natal, Feynman at Princeton University, and the GPC and Niagara supercomputers at the SciNet HPC Consortium \citep{Scinet,Niagara}.
SciNet is funded by the Canada Foundation for Innovation under the auspices of Compute Canada, the Government of Ontario, Ontario Research Fund - Research Excellence, and the University of Toronto.

The collaboration is grateful to the British Antarctic Survey, particularly Sam Burrell, and to the Alfred Wegener Institute and the crew of R.V.  {\it Polarstern} for invaluable assistance with the recovery of the data and payload after the 2015 flight.
Brendan Crill and Tom Montroy made significant contributions to \Spider's development.
This project, like so many others that he founded and supported, owes much to the vision and leadership of the late Professor Andrew E. Lange.

\bibliographystyle{aasjournal}
\bibliography{references}

\end{document}